%% file: GW_Systematics.tex
\documentclass{article}
\pdfoutput=1
\usepackage{jcappub}
\usepackage{amsmath}
\usepackage{amssymb}
\usepackage{hyperref}
\usepackage{multirow}
\usepackage{soul}
\usepackage{bm}
\usepackage{caption}
\usepackage{cancel}

\definecolor{darkred}{RGB}{175,0,0}

\title{Mitigating galaxy systematics with gravitational wave clustering}

\author[a]{Caterina Scarpel,}
\emailAdd{caterina.scarpel@studenti.unipd.it}

\author[a,b,c]{Nicola Bellomo,}
\emailAdd{nicola.bellomo@unipd.it}

\author[d]{Sarah Libanore,}
\emailAdd{libanore@bgu.ac.il}

\author[a,b]{Michele Liguori,}
\emailAdd{michele.liguori@unipd.it}

\author[a,b,c]{Alvise Raccanelli}
\emailAdd{alvise.raccanelli.1@unipd.it}

\affiliation[a]{Dipartimento di Fisica e Astronomia G. Galilei, Università degli Studi di Padova, via Marzolo 8, I-35131 Padova, Italy.}
\affiliation[b]{INFN, Sezione di Padova, Via Marzolo 8, I-35131, Padova, Italy.}
\affiliation[c]{INAF - Osservatorio Astronomico di Padova, Vicolo dell'Osservatorio 5, I-35122 Padova, Italy.}
\affiliation[d]{Department of Physics, Ben-Gurion University of the Negev, Be'er Sheva 84105, Israel.}

\abstract{
Although currently poorly constrained, cosmological ultra-large scales are expected to provide formidable tests not only of General Relativity, but also of the content of~$\Lambda$CDM and the Early Universe.
However, in this regime, cosmic variance plays a major role in limiting sensitivity, and controlling systematic errors becomes a crucial aspect in preserving the limited information content of current and future observations.
Multi-tracer analyses represent a useful technique that simultaneously allows to limit the impact of cosmic variance and mitigate the presence of systematics.
In this sense, gravitational waves might represent the perfect alternative tracer of the large-scale structure, since their detection suffers from a set of uncertainties completely different from that of traditional large-scale structure surveys.
In this work, we provide a concrete example of how gravitational wave clustering mitigates the presence of systematics, and facilitate the discovery of New Physics signatures.
Specifically, we focus on systematics that degrade the constraining power on local Primordial non-Gaussianity for future galaxy surveys; and show how catalogs of gravitational wave events detected by third-generation observatories reduce the impact of systematics while being able to maintain flexibility in the statistical analysis.
Additionally, we release a new version of the~\texttt{Multi\_CLASS} code, which now provides an enhanced level of customization of the different tracers and is compatible with the latest releases of~\texttt{CLASS}.
}

\begin{document}

\maketitle

%%%%%%%%%%%%%%%%%%%%%%%%%%%%%%%%%%%%%%%%%%%%%%%%%%%%%%%%%%%%%%%%%%%%%%%%%%%%%%%%%%%%%%%%%%%%%%%%%%%%%%%%%%%%%%%%%%%%%%%%%%%%%%%%%%%%

\section{Introduction}

Despite being unable to directly probe the Universe at very high energies, Cosmology has already provided a wealth of information about the dynamics of the Early Universe.
In particular, Cosmic Microwave Background observations have been the first experimental probe to strongly hint at the presence of an inflationary epoch that explains the observed homogeneity and isotropy, as well as to provide a dynamical mechanism that generates primordial perturbations on super-horizon scales~\cite{ade:planck2013inflation, ade:planck2015inflation, akrami:planck2018inflation}.
Subsequent observations of the large-scale structure (LSS) of the Universe at late times confirmed and strengthened the inflationary scenario, which today is consistently assumed to be part of the Cosmological Standard Model, the~$\Lambda$CDM.

Unfortunately, the possibility of measuring a limited number of observables over a limited range of cosmological scales severely limits our ability to understand the microphysics of Inflation, which can be accounted for by a plethora of models~\cite{martin:inflationaryencylopedia}, many of which even have good Bayesian evidence~\cite{martin:inflationarymodelsbayesianevidenceI, martin:inflationarymodelsbayesianevidenceII, martin:inflationarymodelsbayesianevidenceIII, martin:inflationarymodelsbayesianevidenceIV}.
One key probe of inflationary dynamics is the presence of Primordial non-Gaussianity (PnG), which carry peculiar imprints created by the processes of that early epoch, see, e.g., Refs.~\cite{bartolo:pngreview, chen:pngreview, renauxpetel:pngreview} and references therein.
In particular, both early-~\cite{komatsu:pngconstraints, ade:pngconstraints, akrami:pngconstraints, jung:pngconstraints} and late-Universe~\cite{damico:lsspngconstraints, cabass:lsspngconstraintsI, rezaie:lsspngconstraints, cabass:lsspngconstraintsII, bermejocliment:lsspngconstraints, chudaykin:lsspngconstraints} experiments have already put impressive constraints on the magnitude and properties of the primordial three-point function of primordial scalar modes, which in the case of perfect Gaussian initial conditions is expected to be exactly zero.

In this work, we focus on a specific class of PnG, those of the local type.
The ultimate sensitivity target in this line of research is set by the guaranteed level of local PnG of single-field, slow-roll models, approximately of the order~$f_\mathrm{NL}\simeq \mathcal{O}(1-n_s) \approx 10^{-2}$~\cite{gangui:ngfrominflation, acquaviva:ngfrominflation, maldacena:ngfrominflation}.
On the other hand, numerous extensions of the standard inflationary scenario allow local PnG with an order-unity magnitude, considerably larger than what single-field slow-roll models predict, see, e.g., Refs.~\cite{enqvist:localpng, lyth:localpng, moroi:localpng, zaldarriaga:localpng}.
Although this ultimate precision is not achievable by galaxy surveys in the near future, we can already start paving the road to that goal, since we expect ongoing surveys to reach a sensitivity of~$\sigma_{f_\mathrm{NL}} \simeq \mathcal{O}( 1)$~\cite{dore:spherexwhitepaperI, dore:spherexwhitepaperII, dore:spherexwhitepaperIII}.
In particular, we want to find a way in which we can mitigate the presence of systematic sources of error.

One potential avenue to increase the constraining power of galaxy survey is by cross-correlating experiments that monitor different LSS tracers, since they provide an ``additional sample'' of the same underlying field~\cite{seljak:cosmicvariance, mcdonald:cosmicvariance}.
Here we argue that a similar argument applies also to the case of increasing the accuracy, because different tracers are inherently subject to different sources of systematic errors.
Although many alternative tracers have already been employed for this ``cross-checking'' analysis, in this work we consider a new one, resolved gravitational wave (GW) events detected by ground-based observatories in the Hz-kHz frequency band.
In particular, we consider a population of binary black holes (BBHs), since they are tightly connected to galaxy evolution and dynamics, thus they also possess an intrinsically anisotropic distribution in the sky.
This connection ensures that GW clustering carries the imprint of the Early Universe dynamics in the form of local PnG.
Additionally, since GW observation is made via interferometers measuring the passage of gravitational radiation and not telescopes/satellites, errors in their detection are completely uncorrelated from those of observations made via electromagnetic radiation.

In this work, we showcase the potential of GWs in mitigating the impact of unknown systematic effects that a galaxy survey can be blind to while testing for the presence of New Physics signatures.
Although poorly localized, GWs provide a unique cross-validation data set to assess the robustness of an eventual PnG detection.
Previous works have already shown the promising potential of GWxLSS cross-correlation in tightening PnG constraints~\cite{gagnon:gwxlsspngconstraints}.
Here we build on that concept to demonstrate how the scope of this methodology is crucial also to increase accuracy by mitigating the presence of additional residual systematics from the data.
In particular, we demonstrate that, almost independently from the type of systematic appearing at large scales, GWxLSS cross-correlations allow for employing a conservative statistical analysis strategy while retaining, if not increasing, the constraining power of the original galaxy survey.
Additionally, we update the existing~\texttt{Multi\_CLASS} code~\cite{bellomo:multiclass, bernal:multiclass} with a new version that is compatible with the structure of~\texttt{CLASS 3.0} and that allows a more realistic characterization of different types of LSS tracers.\footnote{
The new version of the code will be available at~\href{https://github.com/nbellomo/Multi_CLASS}{https://github.com/nbellomo/Multi\_CLASS} after the article has been accepted.}

The paper is organized as follows.
In section~\ref{sec:theoretical_overview} we provide a minimal theoretical overview of the framework implemented in this work.
In section~\ref{sec:impact_systematics} we demonstrate how GWxLSS cross-correlations mitigate the impact of systematics under a variety of circumstances, while in section~\ref{sec:insights_on_systematics} we provide further insights on applications of this technique.
Finally, we conclude in section~\ref{sec:conclusions}.
The appendices~\ref{app:lss_tracers} and~\ref{app:fisher_forecast} contain additional material providing technical details on the implementation and results of this work, while the appendix~\ref{app:multiclass} provides a description of the changes introduced in this new version of~\texttt{Multi\_CLASS}.

%%%%%%%%%%%%%%%%%%%%%%%%%%%%%%%%%%%%%%%%%%%%%%%%%%%%%%%%%%%%%%%%%%%%%%%%%%%%%%%%%%%%%%%%%%%%%%%%%%%%%%%%%%%%%%%%%%%%%%%%%%%%%%%%%%%%

\section{Theoretical overview of the framework}
\label{sec:theoretical_overview}

The backbone of GWxLSS cross-correlations for GW events detected in the Hz-kHz frequency band has already been largely discussed in numerous other works to investigate GW clustering per se, and the nature of dark matter and dark energy, see, e.g., Refs.~\cite{oguri:gwxlss, raccanelli:gwxlssI, raccanelli:gwxlssII, scelfo:gwxlssI, calore:gwxlss, mukherjee:gwxlssI, libanore:gwxlssI, scelfo:gwxlssII, scelfo:gwxlssIII, libanore:gwxlssII, mukherjee:gwxlssII, mukherjee:gwxlssIII, scelfo:gwxlssIV, bosi:gwxlssI, libanore:gwxlssIII, balaudo:gwxlss, afroz:gwxlss, zazzera:gwxlssI, pedrotti:gwxlss, zazzera:gwxlssII, sala:gwxlss, deleo:gwxlss, bosi:gwxlssII}.
Therefore, in this section, we report only the main results necessary to understand the fundamental aspects of this work, and we refer the interested reader to the above set of references for a more general introduction to the topic.

%%%%%%%%%%%%%%%%%%%%%%%%%%%%%%%%%%%%%%%%%%%%%%%%%%%%%%%%%%%%%%%%%%%%%%%%%%%%%%%%%%%%%%%%%%%%%%%%%%%%%%%%%%%%%%%%%%%%%%%%%%%%%%%%%%%%

\subsection{Summary statistics in harmonic space}

The theory underlying the calculation of the number count angular power spectrum was developed in Refs.~\cite{bonvin:numbercountfluctuation, challinor:numbercountfluctuation, jeong:numbercountfluctuation}; however, here we follow the approach of Ref.~\cite{didio:classgal}, which is ultimately the one implemented in~\texttt{CLASS}~\cite{blas:class} and, by extension, in~\texttt{Multi\_CLASS}~\cite{bellomo:multiclass, bernal:multiclass}.
In this formalism, the number density distribution of a tracer~$X$ at redshift~$z$ in the direction~$\hat{\mathbf{n}}$ is decomposed into spherical harmonics as
\begin{equation}
    \delta_{X}(z,\hat{\mathbf{n}}) = \sum_{\ell m} a_{\ell m}^{X,z} Y_{\ell m}(\hat{\mathbf{n}}),
\end{equation}
where~$a_{\ell m}$ are the harmonic coefficients and~$Y_{\ell m}$ are the spherical harmonics.
The angular power spectrum~$C_\ell$ of two generic tracers~$(X,Y)$ at two generic mean redshifts~$(z_i, z_j)$ is given by
\begin{equation}
    \langle a_{\ell m}^{X,z_{i}}a_{\ell'm'}^{Y,z_{j}*} \rangle = \delta_{\ell\ell'}^{K} \delta_{mm'}^{K} C_\ell^{XY}(z_i,z_j),
\end{equation}
where~$\delta^K$ is the Kronecker delta function,
\begin{equation}
    C_\ell^{XY}(z_{i},z_{j}) = 4\pi \int d\log k\ \mathcal{P}_\zeta(k) \Delta_\ell^{X,z_i}(k) \Delta_\ell^{Y,z_j} (k),
    \label{eq:angular_pwr}
\end{equation}
and~$\mathcal{P}_\zeta$ is the almost scale-invariant primordial curvature power spectrum.
The redshift-averaged harmonic transfer functions are defined as
\begin{equation}
    \Delta_{\ell}^{X,z_{i}}(k) = \int_{z_i-\Delta z_i}^{z_i+\Delta z_i} dz \frac{dN_X}{dz} W(z, z_i, \Delta z_i) \Delta_\ell^X(k,z),
\end{equation}
where the harmonic transfer function~$\Delta_{\ell}^{X}$ is given by the sum of density, velocity, lensing, and gravity terms, labeled as in Ref.~\cite{bellomo:multiclass}, $dN_X/dz$ is the number count per redshift bin, $W(z, z_i, \Delta_{z_i})$ is a window function centered in~$z_{i}$ with half-width~$\Delta z_{i}$, such that the product between~$dN_{X}/dz$ and~$W(z,z_{i},\Delta z_{i})$ is normalized to unity in the redshift bin. 

In this work, we consider the capabilities of galaxy and GW clustering measurements in constraining local PnG.
The sensitivity of future experiments is typically assessed by implementing a Fisher matrix analysis~\cite{fisher:fishermatrix, bunn:fishermatrix, vogeley:fishermatrix, tegmark:fishermatrix}.
In this case, the Fisher matrix derived from a Gaussian likelihood in the harmonic coefficients reads as 
\begin{equation}
    F_{\alpha\beta} = \left\langle - \frac{\partial^2 \log\mathcal{L}}{\partial\theta_\alpha \partial\theta_\beta} \right\rangle = \sum_{\ell=2}^{\ell_\mathrm{max}} \frac{2\ell+1}{2} \mathrm{Tr} \left[ \frac{\partial\mathcal{C}_\ell}{\partial\theta_\alpha} \mathcal{C}_\ell^{-1} \frac{\partial\mathcal{C}_\ell}{\partial\theta_\beta} \mathcal{C}_\ell^{-1} \right],
\label{eq:fisher_matrix}
\end{equation}
where~$\ell_\mathrm{max}$ is the maximum multipole of the analysis, $\{\theta_j\}$ are the parameters of interest, $\mathrm{Tr}[\ \cdot\ ]$ is the trace operator, $\mathcal{C}_\ell$ and~$\partial_\theta \mathcal{C}_\ell$ are the covariance matrix and its derivative, respectively.
The elements of the covariance are, as usual, the observed angular power spectra
\begin{equation}
    C_{\ell,\mathrm{obs}}^{XY}(z_i,z_j) = C_\ell^{XY}(z_i,z_j) + N^{XY}_\ell(z_i,z_j),
\end{equation}
where~$N^{XY}_\ell$ is the angular power spectrum of the noise, and it is specified in appendix~\ref{app:lss_tracers} for both galaxies and GWs.
Marginalized errors on the parameters are obtained by inverting the Fisher matrix as~$\sigma^2_{\theta_\alpha} = \left( F^{-1} \right)_{\alpha\alpha}$.

%%%%%%%%%%%%%%%%%%%%%%%%%%%%%%%%%%%%%%%%%%%%%%%%%%%%%%%%%%%%%%%%%%%%%%%%%%%%%%%%%%%%%%%%%%%%%%%%%%%%%%%%%%%%%%%%%%%%%%%%%%%%%%%%%%%%

\subsection{Presence of systematic effects}

Systematics appear in real-life experiments in a variety of forms, from foregrounds to mis-modelling of the observations.
The physics underlying the first category is typically not related to that of the target observations, and it might be possible to create templates of such sources of systematics, also with the help of external experiments.
In this sense, techniques such as template subtraction, either at the map or estimator level, could represent a viable procedure to address the issue, see, e.g., refs.~\cite{ho:templatesubtraction, pullen:templatesubtraction, elsner:templatesubtractionI, kalus:templatesubtractionI, elsner:templatesubtractionII, kalus:templatesubtractionII, weaverdyck:templatesubtraction}.

On the other hand, mis-modelling effects are harder to address with templates, see, e.g., ref.~\cite{awan:rubinsystematic}.
Some very common type of systematics in galaxy surveys are due to imaging, see, e.g., refs.~\cite{kitanidis:imagingsystematics, rezaie:imagingsystematics, chaussidon:imaginsystematics}, and redshift determination, see, e.g., refs.~\cite{rosell:redshiftsystematics, cordero:redshiftsystematics, chan:redshiftsystematics, krolewski:redshiftsystematics}.
Generally speaking, the former category of systematics tends to affect mostly angular modes, while the latter affects radial modes.
Since systematics can appear easily on large scales, their effect might be confused with the presence of a PnG signal~\cite{rezaie:desisystematics, chaussidon:desisystematics},
and numerous works have already analyzed how systematics degrade constraints on this specific extension of~$\Lambda$CDM~\cite{pullen:templatesubtraction, saraf:fnlsystematics, wen:systematicssfb, bruton:fnlsystematics}.

The effect of PnG is present in both classes of modes; thus, in principle galaxy surveys could perform a separate estimate of~$f_\mathrm{NL}$ from radial and angular modes (in the sense of radial/transverse separation).
Alternatively, since different kinds of systematics appear in different modes, it is convenient to adopt some mode decomposition that naturally implements such separation, as done, for instance, in the context of SFB analysis~\cite{liu:systematicssfb, castorina:systematicssfb, wen:systematicssfb, bruton:fnlsystematics} or, as in this paper, for spherical harmonic tomography~\cite{weaverdyck:templatesubtraction}.
Other authors have also implemented strategies that involve the inclusion in the analysis of 1-point functions to detect systematics~\cite{berlfein:systematicmitigation}.
In this work, we explore the alternative path that consists of including in the analysis a different LSS tracer measured by an independent observatory, such as GWs.

In realistic galaxy surveys, the total number of observed objects is given by
\begin{equation}
    N_\mathrm{obs}(\mathbf{x}) = N(\mathbf{x}) \left[ 1 + M(\mathbf{x}) \right] + A(\mathbf{x}) = N(\mathbf{x}) + \delta N_\mathrm{sys}(\mathbf{x}),
\end{equation}
where~$N$ is the intrinsic number of objects, and~$M$ and~$A$ are multiplicative and additive systematics, respectively, where the former contains effects that can typically be encapsulated in the galaxy selection functions, while the latter is typically due to the presence of spurious objects in the galaxy catalog, possibly due to fiber collisions or stellar contamination.
In particular, it is well known that star contamination may lead to the presence of an excess of power at large scales~\cite{shaun:starcontaminationpng, ross:starcontaminationpng}.

Since our goal is not to investigate any specific source of systematics, in the following, we adopt an agnostic approach with respect to the origin of the systematic effect.
In other words, our reasoning is not limited to a specific scenario, but it can also be broadly applied to the presence of multiple types of systematics, even in the case of ``unknown unknowns''.
Following this spirit, we argue that the additional number count fluctuation~$\delta N_\mathrm{sys}(\mathbf{x})$ generated by systematics effectively changes the clustering strength of the galaxy population; therefore, it might be possible to reabsorb its effects into a ``systematic bias'' parameter~$b_\mathrm{sys}$ in such a way that the total galaxy bias now reads as~$b_\mathrm{gal} = b_\mathrm{phys} + \delta b_\mathrm{sys}$.
This is not surprising, especially for the case where multiplicative systematics are present.

In the case of local PnG, the physical bias is given by~\cite{dalal:nghalobias, matarrese:nghalobias, slosar:nghalobias, afshordi:nghalobias, mcdonald:nghalobias}
\begin{equation}
    b_\mathrm{phys}(k,z) = b_\mathrm{G}(z) + b_\mathrm{NG}(k,z) = b_\mathrm{G}(z) + b_\phi f_\mathrm{NL} \mathcal{T}^{-1}_m (k,z),
\end{equation}
where~$b_\mathrm{G}$ and~$b_\mathrm{NG}$ are the Gaussian and non-Gaussian components of the bias, $b_\phi$ is the assembly bias~\cite{gao:assemblybias}, and~$\mathcal{T}_m$ is the matter transfer function.
In general, the assembly bias is given by the sum of a universal component and a second one that depends on the halo formation history, and reads as~\cite{slosar:nghalobias, reid:assemblybias}
\begin{equation}
    b_\phi = b^\mathrm{univ}_\phi + \Delta b_\phi = 2\delta_c (b_G-1) + 2\delta_c (1-p),
\end{equation}
where~$\delta_c$ is the linearly extrapolated overdensity for gravitational collapse, and~$p$ is an effective parameter that characterizes different tracers.
Therefore, at large scales, we expect the total bias to be dominated by the non-Gaussian contribution and scale as~$b_\mathrm{phys} \propto k^{-2}$.
Therefore, since our interest is in systematics that limit the constraining power of PnG, in this work we parametrize the systematic bias as
\begin{equation}
    \delta b_\mathrm{sys} = b_\mathrm{sys} \left( \frac{k}{k_\mathrm{sys}} \right)^{\alpha_\mathrm{sys}},
\end{equation}
where~$\{b_\mathrm{sys}, \alpha_\mathrm{sys}, k_\mathrm{sys}\}$ are a amplitude, tilt, and pivot scale parameter, respectively.
The pivot scale is totally degenerate with the amplitude; thus, we conveniently fix it to the pivot scale of the primordial curvature power spectrum, i.e., ~$k_\mathrm{sys}=0.05\ \mathrm{Mpc}^{-1}$.
Regarding the tilt, we are mainly interested in the~$\{ \alpha_\mathrm{sys} = -2 \}$ scenario, since it will be highly degenerate with the signature generated by local PnG.
However, we also investigate the~$\{ \alpha_\mathrm{sys} = -1,-3 \}$ scenarios to better grasp the effectiveness of cross-correlations in disentangling systematic effects.

Since we are working in a phenomenological scenario, no a priori value of the systematic bias parameter is available.
On the other hand, choosing an arbitrary value of~$b_\mathrm{sys}$ might end up overestimating the constraining power of cross-correlations.
Therefore, in the following, we develop a strategy to identify the most conservative choice of~$b_\mathrm{sys}$.
The idea is to scan a range of values for this parameter with the goal of minimizing the function
\begin{equation}
    \mathcal{D}(b_\mathrm{sys}) = f_\mathrm{sky }\sum_{\ell=2}^{\ell_\mathrm{max}} \frac{2\ell+1}{2} \mathrm{Tr} \left[ \left( \mathcal{C}^\mathrm{sys}_\ell - \mathcal{C}^\mathrm{PnG}_\ell \right) \left( \mathcal{C}^\mathrm{PnG}_\ell \right)^{-1} \left( \mathcal{C}^\mathrm{sys}_\ell - \mathcal{C}^\mathrm{PnG}_\ell \right) \left( \mathcal{C}^\mathrm{PnG}_\ell \right)^{-1} \right],
\label{eq:bsys_distinguishability}
\end{equation}
which attempts to measure the distinguishability of a model with systematics, but zero PnG, from a fiducial model with PnG, but zero systematics.
Additionally, if in the minimum of this function we have~$\mathcal{D}(b^\mathrm{min}_\mathrm{sys}) \lesssim \sigma^2_{f_\mathrm{NL}}$, we are in the conservative scenario where the two models cannot be disentangled using only the galaxy data set.
On the other hand, if for any choice of the exponent~$\alpha_\mathrm{sys}$ we have~$\mathcal{D}(b^\mathrm{min}_\mathrm{sys}) \gtrsim \sigma^2_{f_\mathrm{NL}}$, we can safely conclude that the presence of a systematic error is detectable by the galaxy survey alone, without the need of employing cross-correlations.

%%%%%%%%%%%%%%%%%%%%%%%%%%%%%%%%%%%%%%%%%%%%%%%%%%%%%%%%%%%%%%%%%%%%%%%%%%%%%%%%%%%%%%%%%%%%%%%%%%%%%%%%%%%%%%%%%%%%%%%%%%%%%%%%%%%%

\subsection{Tracers of large-scale structure}

Galaxies and GWs trace in similar, yet not equivalent, fashions the same underlying dark matter field.
In terms of number count anisotropy, this difference translates into specific sets of bias functions appearing at the linear level in perturbation theory, i.e., clustering, assembly (if local PnG are present), magnification, and evolution bias.
Their observations are also characterized by radically different uncertainties, for instance in terms of radial sensitivity, requiring a dedicated window function treatment.
The new version of~\texttt{Multi\_CLASS} that accompanies this article is born to address all these peculiarities.\footnote{
The new version of the code will be available at~\href{https://github.com/nbellomo/Multi_CLASS}{https://github.com/nbellomo/Multi\_CLASS} after the article has been accepted.}
In addition to being compatible with the new architecture of~\texttt{CLASS v3.0} and subsequent releases, this new version allows the user to specify for each individual tracer its set of binning choices, window functions, and bias functions.
Moreover, in this new version, the value of the bias functions is no longer passed as a ``bin-averaged'' value, but as a time-dependent quantity, removing a~$\mathcal{O}(1-10\%)$ source of systematic uncertainty from the computation of the angular power spectra.

Since we observe these different tracers in totally independent fashions, they represent each other perfect counterpart to assess not only the consistency of cosmological evolution, but also the robustness of the very same cosmological analysis.
Since our goal is to show the interplay between next-generation galaxy surveys and GW observatories, we consider
\begin{itemize}
    \item a \textbf{SPHEREx-like galaxy survey} in terms of the observed population of galaxy, survey depth, binning strategy, and sky coverage.
    The complete characterization of this galaxy population is reported in appendix~\ref{subapp:galaxy_sample}.

    \item a collection of \textbf{GW events detected by a third-generation network}, specifically Einstein Telescope and two Cosmic Explorers.
    The source of these GWs is a population of astrophysical BBHs with properties compatible with current observations.
    Although well constrained in the low-redshift Universe, the merger rate still represents a considerable source of uncertainty.
    Therefore, the total number of observed events that future interferometers will observe is currently unknown.
    In this work, we consider three scenarios called ``conservative'', ``optimistic'', and ``futuristic'', corresponding to a number of GW events observed per year of~$N_\mathrm{BH}=10^5, 10^6, 3\times10^6\ \mathrm{GW/yr}$, respectively.
    The value of the conservative scenario represents the current baseline used by the Einstein Telescope collaboration~\cite{abac:etscience}.
    The total observation time is set to~$T_\mathrm{obs}=10\ \mathrm{yr}$ in all scenarios.
    
    The localization of GW events will be known very poorly in comparison to that of galaxies, even for next generation observatories.
    In this analysis, we consider a median angular resolution of~$\Delta\Omega \approx 1.5\ \mathrm{deg}^2$, and accordingly restrict the GW clustering analysis to a maximum multipole of~$\ell_\mathrm{GW}^\mathrm{res}=100$~\cite{bosi:gwxlssI}.
    This localization resolution defines our separation between the ``large'' and ``small'' scales, as we explain in the next section.
    The complete characterization of this GW population is reported in the appendix~\ref{subapp:gw_sample}.
\end{itemize}

Finally, we stress that the specifics of each tracer have not been designed to maximize the scientific potential of the case at hand, but to be representative of future realistic datasets.
Therefore, there is plenty of room for improvements in terms of constraining power on cosmological, bias, New Physics, and systematic parameters.
We leave the investigation of what represents the ideal optimization strategy for each of these analyses for future work.

%%%%%%%%%%%%%%%%%%%%%%%%%%%%%%%%%%%%%%%%%%%%%%%%%%%%%%%%%%%%%%%%%%%%%%%%%%%%%%%%%%%%%%%%%%%%%%%%%%%%%%%%%%%%%%%%%%%%%%%%%%%%%%%%%%%%

\section{Impact of systematics}
\label{sec:impact_systematics}

We introduce a set of likelihoods describing different steps of a multi-tracer statistical analysis to better understand the interplay between physical effects, systematics, and different tracers.
Specific attention is devoted to the large scale analysis, since the signatures of local PnG are more pronounced in this regime.
Therefore, we consider for the galaxy survey
\begin{equation}
    \mathcal{L}_\mathrm{gal, LS}, \qquad \qquad \mathcal{L}^\mathrm{LSS}_\mathrm{tot} = \mathcal{L}_\mathrm{gal, LS} \times \mathcal{L}_\mathrm{gal, SS},
\label{eq:galaxy_likelihood}
\end{equation}
where~$\mathcal{L}_\mathrm{gal, LS}$ and~$\mathcal{L}_\mathrm{gal, SS}$ are the likelihoods at large and small scales, respectively.
For the purpose of this work, ``large scales'' correspond to the range of multipoles~$\ell\in[2, 100]$, where a multi-tracer analysis can be meaningful; whereas ``small scales'' include multipoles~$\ell>100$ where linear perturbation theory still applies, as explained in appendix~\ref{app:lss_tracers}, but GW clustering signal is heavily suppressed.
Therefore, once we incorporate GW clustering into the analysis, the total likelihood is given by
\begin{equation}
    \mathcal{L}^\mathrm{GWxLSS}_\mathrm{tot} = \mathcal{L}_\mathrm{gal/GW, LS} \times \mathcal{L}_\mathrm{gal, SS},
\label{eq:gwxlss_likelihood}
\end{equation}
where~$\mathcal{L}_\mathrm{gal/GW, LS}$ is the galaxy-GW joint-likelihood at large scales only.

In each scenario, the Fisher matrix is computed using the definition provided in equation~\eqref{eq:fisher_matrix}.
The parameters varied in the Fisher matrix analysis are the cold dark matter physical density~$\omega_\mathrm{cdm}$, the Hubble expansion rate today~$h$, the primordial power spectrum spectral tilt~$n_s$, the sets of bias parameters~$\{ b_{j,\mathrm{gal}} \}$ and~$\{ b_{j,\mathrm{GW}} \}$ describing the galaxy and GW populations, respectively, the local PnG parameter~$f_\mathrm{NL}$, and the systematic bias amplitude~$b_\mathrm{sys}$.
The presence of systematics is considered exclusively for the galaxy population.
The fiducial values for the baseline fiducial model of this work, called ``PNG'', are reported in table~\ref{tab:fisher_fiducial_values} of appendix~\ref{app:fisher_forecast}.
We assume a universal assembly bias, i.e.,~$\Delta b_\phi=0$ or~$p=1$; however, in section~\ref{sec:insights_on_systematics} we comment about different scenarios where this assumption is not taken.

%%%%%%%%%%%%%%%%%%%%%%%%%%%%%%%%%%%%%%%%%%%%%%%%%%%%%%%%%%%%%%%%%%%%%%%%%%%%%%%%%%%%%%%%%%%%%%%%%%%%%%%%%%%%%%%%%%%%%%%%%%%%%%%%%%%%

\subsection{Baseline sensitivity}

First, we establish the baseline sensitivity of the galaxy survey configuration to constrain cosmological, bias, and, in particular, PnG parameters.
At this stage, we do not consider the possibility of the existence of systematics, i.e., the systematic bias amplitude is not included in the set of parameters entering the Fisher matrix analysis.
We refer to this subclass of the analysis as~``PNG/no-sys'', and we report the marginalized errors in table~\ref{tab:PNG_errors} of appendix~\ref{app:fisher_forecast} for the LS only, SS only, and LS+SS cases.

\begin{figure}[ht]
    \centerline{
    \includegraphics[width=\linewidth]{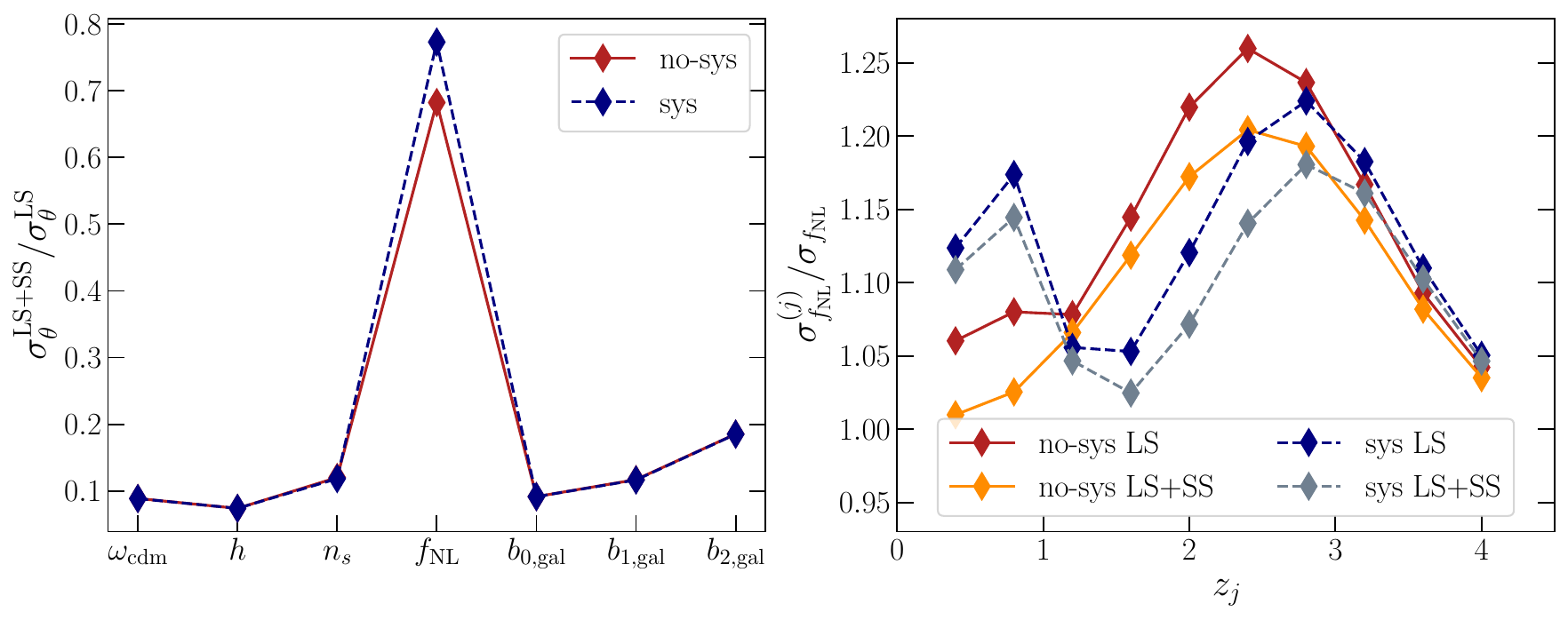}}
    \caption{\textit{Left panel}: marginalized error ratio between the total and large-scale only analysis, for both PNG/no-sys (\textit{red solid line}) and PNG/sys (\textit{blue dashed line}) subclasses of the analysis. 
    \textit{Right panel}: ratio of marginalized errors with and without the exclusion of one individual redshift bin from the analysis, and with and without the inclusion of small scales.
    \textit{Solid} and \textit{dashed lines} refer to the PNG/no-sys and PNG/sys subclasses of the analysis, respectively.}
    \label{fig:PNGerrors_and_dropoutanalysis}
\end{figure}

Although these numbers might not be new per se, they provide a useful benchmark to test the impact of systematics.
The solid red line in the left panel of figure~\ref{fig:PNGerrors_and_dropoutanalysis} describes the ratio of marginalized error with and without the inclusion of small scales in the forecast.
As expected, we note an overall improvement in the constraining power on both cosmological and bias parameters by almost one order of magnitude; however, in the case of~$f_\mathrm{NL}$, the gain is only of about~$30\%$.
The gain in constraining power for PnG is not driven by extra information at~$\ell \gtrsim 100$, since at these scales the PnG contribution is subdominant, but by breaking the degeneracies between~$f_\mathrm{NL}$ and the other cosmological and bias parameters.

Additionally, for the purpose of clarifying where systematics might play the greatest role, we need to characterize which redshift bin has the largest constraining power in terms of~$f_\mathrm{NL}$. 
However, since information on PnG is contained not only in auto-bin, but also in the cross-bin angular power spectra because of density-lensing terms, it is not possible to directly analyze the constraining power coming from individual bins.
Thus, we resort to a dropout analysis, in which we compare the total error~$\sigma_{f_\mathrm{NL}}$, obtained from the analysis where all redshift bins are included, with the error~$\sigma^{(j)}_{f_\mathrm{NL}}$ obtained by excluding a single redshift bin centered at~$z_j$ from the analysis.
In other words, we perform the same kind of analysis of equation~\eqref{eq:fisher_matrix} where the dimension of the covariance matrices is one less than in the standard case.
We show the result of the dropout analysis with solid lines in the right panel of figure~\ref{fig:PNGerrors_and_dropoutanalysis}, with and without the inclusions of small-scales. 
As we observe, different competing effects play a role in determining where the real constraining power comes from. 
On the one hand, low-redshift bins are the least noise-dominated; on the other hand, at high-redshift the effect of local PnG is more pronounced, since~$b_\mathrm{NG}/b_\mathrm{G} \propto (1+z)$.
In the end, in this setup, redshift bins in the intermediate range~$2\lesssim z\lesssim3$ strike some balance between these two effects and appear to contain the largest information content.

%%%%%%%%%%%%%%%%%%%%%%%%%%%%%%%%%%%%%%%%%%%%%%%%%%%%%%%%%%%%%%%%%%%%%%%%%%%%%%%%%%%%%%%%%%%%%%%%%%%%%%%%%%%%%%%%%%%%%%%%%%%%%%%%%%%%

\subsection{Mitigation of systematics with GWxLSS}

Once the baseline analysis has been established, we introduce the possibility of exploring the presence of systematics in the analysis of the galaxy population.
In other words, in this case, we are asking the data how easily the PnG features can be explained by a systematic template.
At the forecast level, this is equivalent to the addition of the systematics bias amplitude to the set of cosmological and bias parameters, keeping the GW population free from any systematics.
We call this subclass of the analysis ``PNG/sys'', and we report marginalized errors obtained in this setting in table~\ref{tab:PNG_errors} of the appendix~\ref{app:fisher_forecast} for the large-scale only, small-scale only, and total cases.
The dashed lines in figure~\ref{fig:PNGerrors_and_dropoutanalysis} illustrate the improvement in marginalized errors due to small scales, and the redshift bins that host most of the constraining power.
Adding the systematic bias parameter does not qualitatively change the conclusions of the previous section.

\begin{figure}[ht]
    \centerline{
    \includegraphics[width=\linewidth]{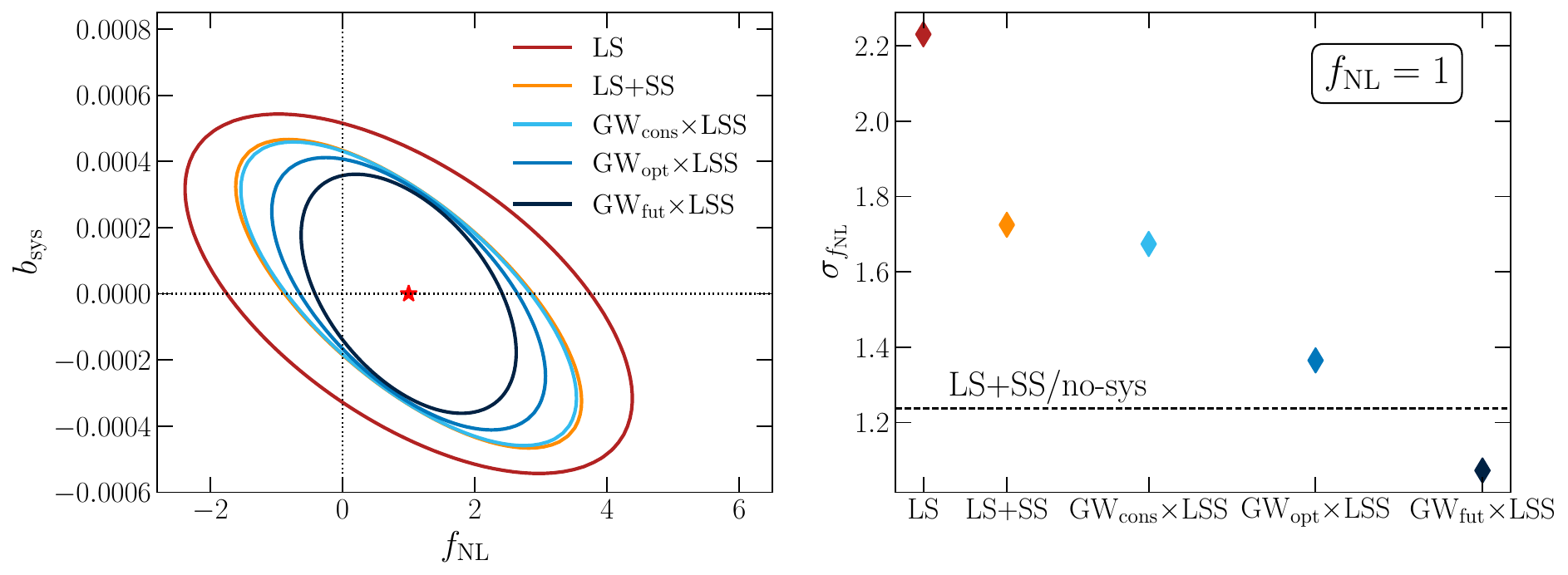}}
    \caption{\textit{Left panel}: two-dimensional marginalized~$68\%$ confidence regions for~$f_\mathrm{NL}$ and~$b_\mathrm{sys}$ for galaxy-only (\textit{red} and \textit{orange} lines) and GWxLSS (\textit{light blue}, \textit{blue} and \textit{dark blue} lines) PNG/sys analysis.
    The \textit{red star} marks the position of the PNG fiducial model.
    \textit{Right panel}: marginalized error on~$f_\mathrm{NL}$ from galaxy-only and GWxLSS PNG/sys analysis. 
    The dashed black line represents the value of the PNG/no-sys galaxy-only analysis.}
    \label{fig:bsys_fnl_degeneracy}
\end{figure}

We show in the left panel of figure~\ref{fig:bsys_fnl_degeneracy} the expected degeneracy between systematics and New Physics parameters.
As immediately appears, scale-dependent systematics severely impact the constraining power of LSS surveys; however, thanks to the inclusion of additional tracers in the analysis, the constraining power is gradually recovered.
The overall improvement depends on the number of available GW events, and it provides constraints on PnG that are approximately~$25-50\%$ and~$5-40\%$ tighter than in the LS and LS+SS galaxy-only analysis, respectively.
The right panel of figure~\ref{fig:bsys_fnl_degeneracy} compares the systematic-free analysis of the previous section with the scenario where systematics are allowed to be present.
Also in this case, we show how the presence of systematics substantially degrades the constraining power - especially for the LS only case - and how GWs gradually restore and even exceed the original constraining power on~$f_\mathrm{NL}$ for~$N_\mathrm{BH}\gtrsim 20\times 10^6$, while providing a greater degree of flexibility in the analysis.

\begin{figure}[ht]
    \centerline{
    \includegraphics[width=\linewidth]{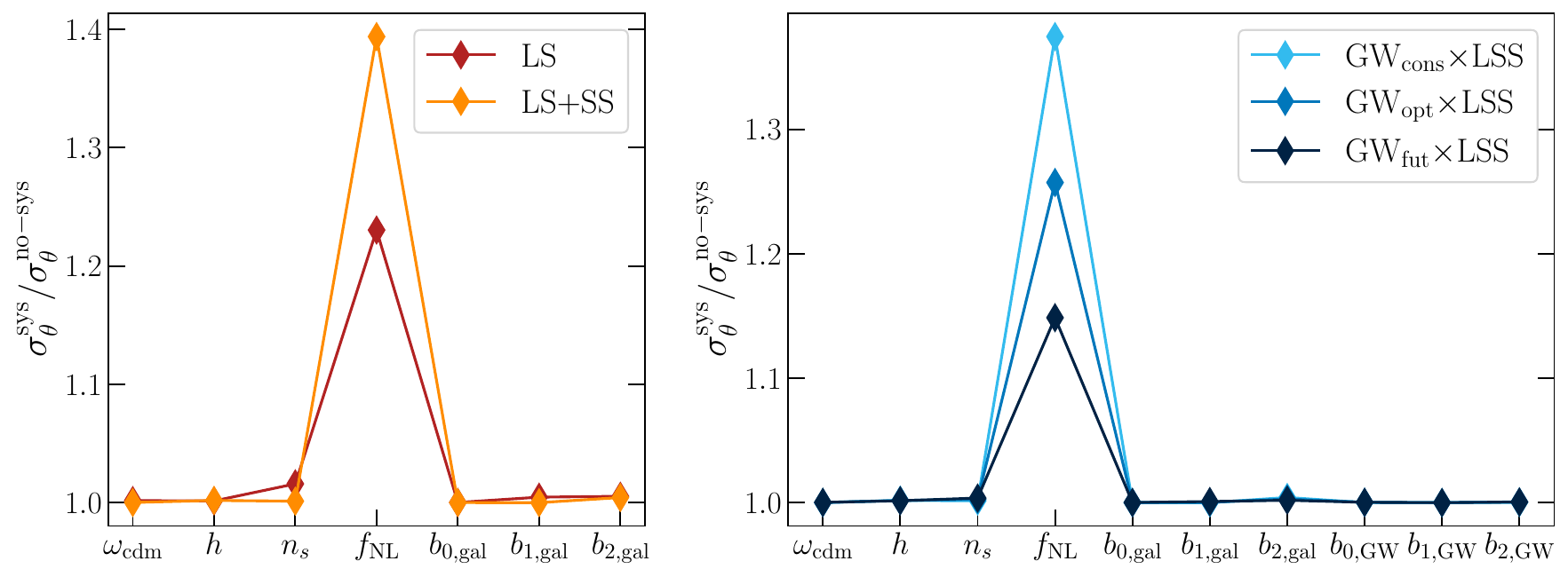}}
    \caption{Ratio of marginalized errors between the PNG/sys and PNG/no-sys analyses, i.e., with and without the inclusion of the systematic bias parameter in the Fisher Matrix, for galaxy-only (\textit{left panel}) and GWxLSS (\textit{right panel}).}
    \label{fig:PNGsys_vs_PNGnosys}
\end{figure}

The importance of performing a multi-tracer analysis with independent data sets is made even more clear by figure~\ref{fig:PNGsys_vs_PNGnosys}.
This figure compares the marginalized errors of the PNG/sys and PNG/no-sys analyses, and it shows how scale-dependent systematics are fundamentally degenerate only with PnG.
In other words, the presence of systematics would not have any significant impact on other parameters constraining power, further hiding their presence.
Therefore, only the presence of an independent tracer that does not present that kind of systematic is able to constrain, or possibly even reveal, their existence.
This result holds independently of whether or not GWs are included.

%%%%%%%%%%%%%%%%%%%%%%%%%%%%%%%%%%%%%%%%%%%%%%%%%%%%%%%%%%%%%%%%%%%%%%%%%%%%%%%%%%%%%%%%%%%%%%%%%%%%%%%%%%%%%%%%%%%%%%%%%%%%%%%%%%%%

\subsubsection{Sensitivity to the assumptions of the fiducial model}

In this section, we test how strongly the systematic error mitigation strategy depends on certain assumptions implicitly made in the choice of the fiducial model.
For instance, the choice of a parametric form for the systematic effect has been motivated by the study of generic behavior; however, it is reasonable to wonder how these results generalize when the setup is altered.
Therefore, in the following, we investigate the robustness of our general message by varying \textit{(i)} the systematic exponent~$\alpha_\mathrm{sys}$, \textit{(ii)} the GW binning scheme, and~\textit{(iii)} the fiducial value of~$f_\mathrm{NL}$.

\begin{figure}[ht]
    \centerline{
    \includegraphics[width=\linewidth]{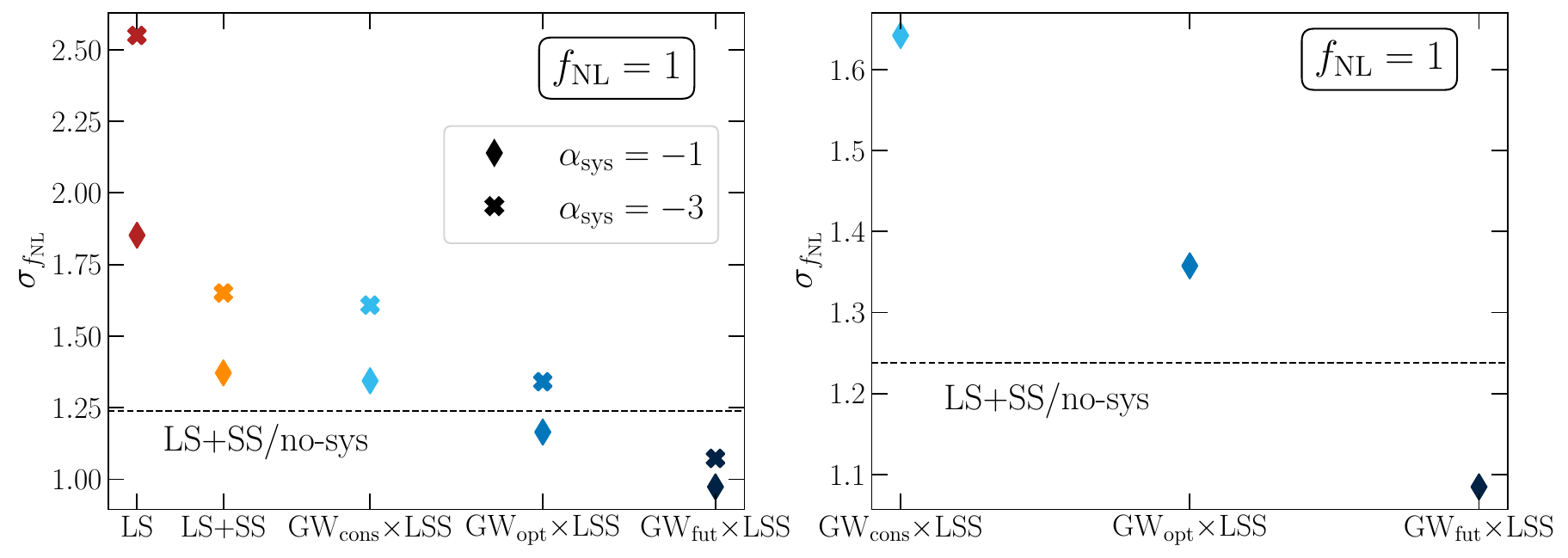}}
    \caption{\textit{Left panel}: marginalized error on PnG for the~$\alpha_\mathrm{sys}=-1, -3$ from galaxy-only and GWxLSS PNG-sys analyses. 
    The dashed black line represents the value of the PNG/no-sys analysis.
    \textit{Right panel}: marginalized error on~$f_\mathrm{NL}$ from GW$\times$LSS PNG/sys analysis with GW events organized into 4 redshift bins.
    The dashed black line represents the value of the PNG/no-sys analysis.}
    \label{fig:asys_4bins}
\end{figure}

Different systematic effects might present different scale dependence.
In particular, stellar templates suggest that star contamination is responsible for a systematic signal that scales as~$(\ell+1)^{-2.3}$~\cite{wen:systematicssfb}, which approximately corresponds to the~$\alpha_\mathrm{sys}=-1$ scenario.
We show in the left panel of figure~\ref{fig:asys_4bins} the marginalized errors on~$f_\mathrm{NL}$ for different values of systematic tilt~$\alpha_\mathrm{sys}$.
The trend displayed in the panel follows that of the case with~$\alpha_\mathrm{sys}=-2$; however, the degeneracy between PnG and systematic effects with small exponents becomes harder to disentangle, since the range of scales where both contributions are present is limited.
Nevertheless, GWs - especially when their total number exceeds ten million - are still able to bring uncertainties under control and even improve on the galaxy-only analysis.
Additionally, as we show in the left panel of figure~\ref{fig:asys_4bins_params} in the appendix~\ref{app:fisher_forecast}, for values of~$\alpha_\mathrm{sys}\gtrsim -2$, we start observing some level of degeneracy between systematics and cosmological/bias parameters.
This finding is easily explained: the ``flatter'' the systematic effect is, the larger its degeneracy is with the small tilt of the primordial power spectrum, or with the overall amplitude of the angular power spectrum, which is regulated by the bias of the tracers and the amount of cold dark matter.
On the other hand, this effect does not appear to be present for~$\alpha_\mathrm{sys}\lesssim -2$, since nothing on linear scales produces any feature with such characteristic scale dependence.

Second, we consider the impact of a different GW binning scheme.
As already stated, the goal of this work is not finding the optimal binning scheme to constrain PnG; however, this type of check proves to be extremely useful in testing the generality of the reported results.
In particular, here we want to test whether it is more effective in terms of constraining power to decrease the shot noise by making larger bins with more GW events, or to increase the number of the measured cross-correlations angular power spectra by making thinner bins.
In this case, we join nearby redshift bins, obtaining a new set of mean redshift and half-width for the GW population given by~$z_j=\{ 0.6, 1.4, 2.2, 3.3 \}$ and~$\Delta z = \{ 0.4, 0.4, 0.4, 0.7 \}$.
Since now the uncertainty of each GW event is smaller than the bin width, we choose a top-hat window function.
We report the results of our analysis in terms of~$f_\mathrm{NL}$ in the right panel of figure~\ref{fig:asys_4bins}, and in terms of the other cosmological parameter in the right panel of figure~\ref{fig:asys_4bins_params} of the appendix~\ref{app:fisher_forecast}. 
In summary, we find that the errors on the cosmological and galaxy bias parameters are basically unaffected by the rebinning, and that errors on~$f_\mathrm{NL}$ are further lowered by a few percent in the case of a large number of GW events.
In this sense, it appears that the GW rebinning does not impact our ability to measure PnG.
However, we also observe that the errors on the GW bias parameters worsen by~$10-20\%$ with respect to the analysis done with eight redshift bins.
In other words, rebinning might play a role in all those studies whether an accurate measure of GW bias is crucial in extracting information about binary formation mechanisms.
This simple rebinning strategy signals the importance of systematically looking for the optimal interplay between the number of measured cross-correlation angular power spectra and the role of shot-noise.
For the moment, a fully comprehensive analysis of this problem is left for future work.

\begin{figure}[ht]
    \centerline{
    \includegraphics[width=\linewidth]{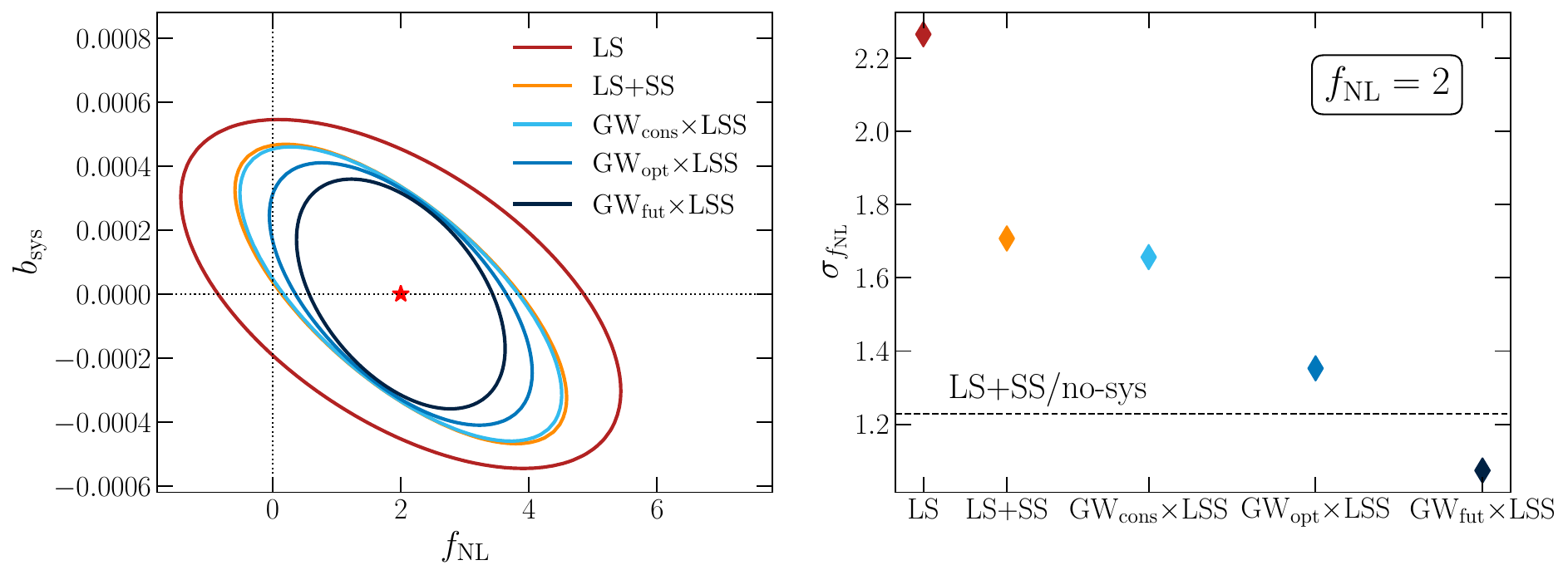}}
    \caption{\textit{Left panel}: two-dimensional marginalized~$68\%$ confidence regions for~$f_\mathrm{NL}$ and~$b_\mathrm{sys}$ for galaxy-only (\textit{red} and \textit{orange} lines) and GWxLSS (\textit{light blue}, \textit{blue} and \textit{dark blue} lines) PNG/sys analysis.
    The \textit{red star} marks the position of the new fiducial model with~$f_\mathrm{NL}=2$.
    \textit{Right panel}: marginalized error on~$f_\mathrm{NL}$ from galaxy-only and galaxy-GW PNG/sys analysis. 
    The dashed black line represents the value of the PNG/no-sys galaxy-only analysis with~$f_\mathrm{NL}=2$.}
    \label{fig:bsys_fnl2_degeneracy}
\end{figure}

Finally, we investigate the case where the Universe is characterized by local PnG with fiducial value~$f_\mathrm{NL}=2$.
We find that in this case the overall sensitivity is very similar to the~$f_\mathrm{NL}=1$ scenario in terms of cosmological, bias, PnG, and systematics parameters.
As for the previous cases, GWs turn out to be extremely effective in reducing the impact of systematics, as we observe in figure~\ref{fig:bsys_fnl2_degeneracy}.
However, in the case of PnG with~$f_\mathrm{NL}\gtrsim 2$, we also observe that, thanks to GWxLSS cross-correlations, we recover enough constraining power to start ruling out values of~$f_\mathrm{NL} \lesssim 0.1$ at almost~$2\sigma$ level, i.e., the range of PnG that contains the predicted value of~$f_\mathrm{NL}$ from single-field slow-roll inflationary scenarios.
In other words, despite introducing a larger degree flexibility in the analysis compared to the PNG/no-sys case, cross-correlations preserve the possibility of ruling out the simplest set of models that explain the first fraction of seconds of the life of our Universe.

%%%%%%%%%%%%%%%%%%%%%%%%%%%%%%%%%%%%%%%%%%%%%%%%%%%%%%%%%%%%%%%%%%%%%%%%%%%%%%%%%%%%%%%%%%%%%%%%%%%%%%%%%%%%%%%%%%%%%%%%%%%%%%%%%%%%

\subsection{Systematics in the data}

The cross-correlation strategy also allows us to address a complementary kind of situation in which we investigate the detection of PnG in the case where systematics are \textit{known} to be present in the data.
In other words, this time our fiducial model, called ``SYS'' and specified in table~\ref{tab:fisher_fiducial_values} of the appendix~\ref{app:fisher_forecast}, has non-zero systematics, but zero PnG.

\begin{figure}[ht]
    \centerline{
    \includegraphics[width=\linewidth]{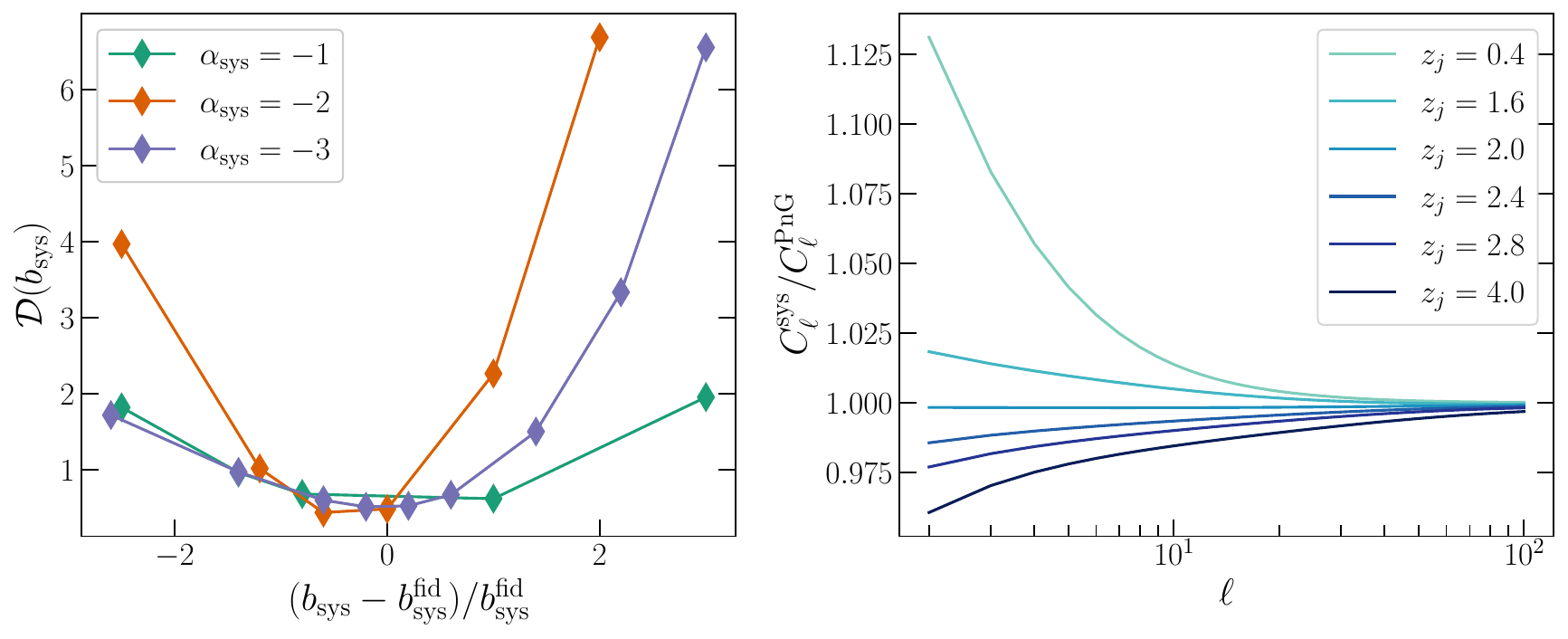}}
    \caption{\textit{Left panel}: distinguishability of different SYS fiducial models compared to the PNG one for different values of the systematic tilt parameter.
    \textit{Right panel}: ratio of angular power spectra corresponding to the PNG and SYS $(\alpha_\mathrm{sys}=-2)$ fiducial models in different redshift bins.}
    \label{fig:optimalbias}
\end{figure}

The value of~$b_\mathrm{sys}$ in the SYS fiducial model is selected to be ``maximally degenerate'' with the PNG fiducial model.
In this sense, the results of this section are to be understood as conservative, since we are considering the worst case scenario allowed by our parametrization.
The degree of degeneracy is assessed using equation~\eqref{eq:bsys_distinguishability}, and is shown in the left panel of figure~\ref{fig:optimalbias} for different values of~$\alpha_\mathrm{sys}$.
Since the minimum of each curve is below the threshold value of~$\sigma^2_{f_\mathrm{NL}}$, we are always in the situation where the PNG and SYS fiducial models are statistically indistinguishable with a galaxy-only analysis.
In the right panel of the same figure, we observe that equation~\eqref{eq:bsys_distinguishability} effectively tries to minimize the differences between fiducial models in low- and high-redshift angular power spectra, also in light of the information content of each bin, as also discussed in the baseline analysis section.
We choose as fiducial values of the systematic bias amplitude~$b_\mathrm{sys}= \{1.0\times 10^{-3}, 2.0\times 10^{-4}, 2.5\times 10^{-6}\}$ for~$\alpha_\mathrm{sys} = \{-1, -2, -3 \}$, respectively.

\begin{figure}[ht]
    \centerline{
    \includegraphics[width=\linewidth]{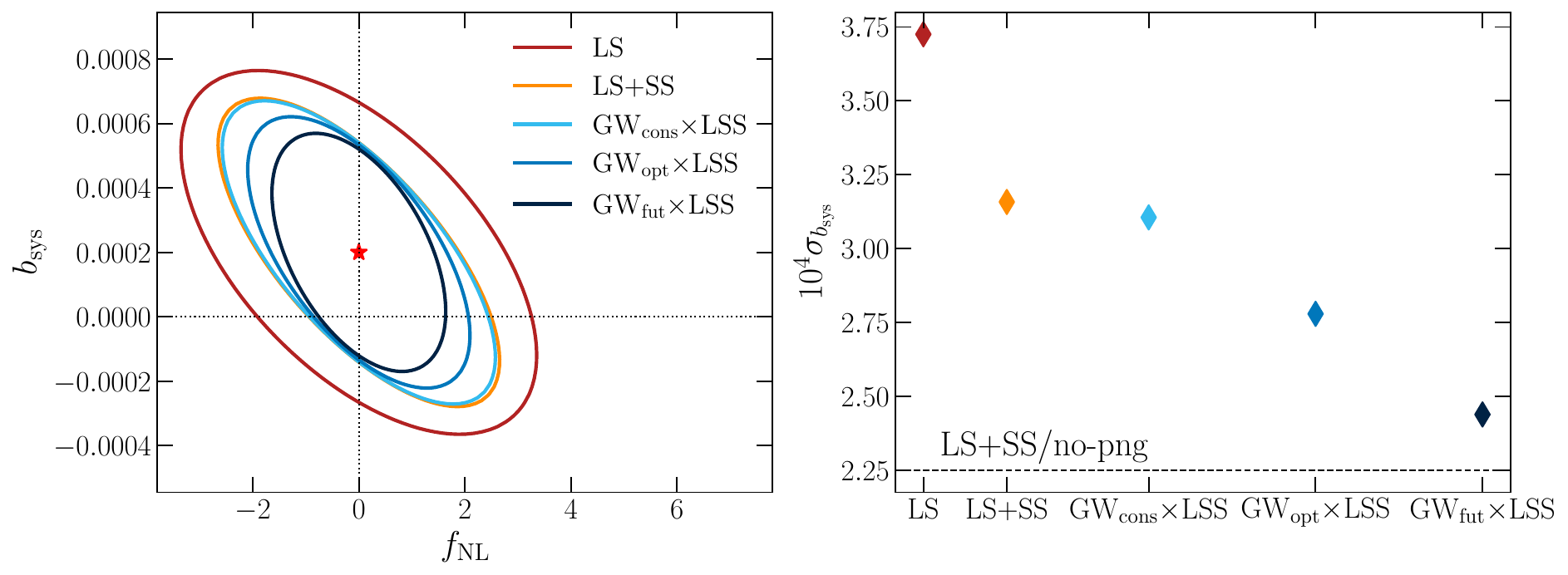}}
    \caption{\textit{Left panel}: two-dimensional marginalized~$68\%$ confidence regions for~$f_\mathrm{NL}$ and~$b_\mathrm{sys}$ for galaxy-only (\textit{red} and \textit{orange} lines) and GWxLSS (\textit{light blue}, \textit{blue} and \textit{dark blue} lines) SYS/png analysis.
    The \textit{red star} marks the position of the SYS fiducial model.
    \textit{Right panel}: marginalized error on~$b_\mathrm{sys}$ from galaxy-only and GWxLSS SYS/png analysis. 
    The dashed black line represents the value of the SYS/no-png galaxy-only analysis.}
    \label{fig:bsys_fnl_degeneracy_SYS}
\end{figure}

As before, we perform two subclasses of analysis, SYS/no-png and SYS/png, where we exclude and include in the Fisher Matrix the local PnG parameter~$f_\mathrm{NL}$, respectively.
The results of the Fisher forecasts for both analyses are reported in table~\ref{tab:SYS_errors} of the appendix~\ref{app:fisher_forecast}.
We show in figure~\ref{fig:bsys_fnl_degeneracy_SYS} the expected degeneracy between systematics and PnG parameters, along with the marginalized errors in the systematics bias amplitude.
As before, we observe the effectiveness of GWxLSS cross-correlation in mitigating this degeneracy.
In particular, we see that with approximately ten million GW events we can start ruling out models of PnG with~$f_\mathrm{NL}\gtrsim 2$, even in the presence of known systematics in our galaxy data.
However, as shown in the right panel of the same figure, in the SYS/png case we are never able to recover the same constraining power on~$b_\mathrm{sys}$ that we have in the SYS/no-png analysis, at least for the number of GW events considered in this analysis.
This result highlights the importance of testing fiducial models in different setups when exploring theoretical alternative explanations, since the results of forecasts are intrinsically dependent on the chosen fiducial.

Finally, even if not explicitly reported for the sake of conciseness, we tested the robustness of our results as done in the previous section.
First, we find that the constraining power in terms of the cosmological and bias parameters is within~$10\%$ of the reported values for the PNG fiducial model.
The inclusion of galaxy small-scales yields an improvement very similar to what reported in the left panel of figure~\ref{fig:PNGerrors_and_dropoutanalysis}.
The addition of the GW dataset helps break the degeneracies and tighten the constraints on all the parameters by a similar amount to what is reported in the previous case, as can be appreciated by comparing tables~\ref{tab:SYS_errors} and~\ref{tab:PNG_errors}.
Finally, varying the systematic bias tilt does not affect the qualitative aspect of the above discussion, it just changes the absolute value of the marginalized errors by no more than~$10-20\%$.

%%%%%%%%%%%%%%%%%%%%%%%%%%%%%%%%%%%%%%%%%%%%%%%%%%%%%%%%%%%%%%%%%%%%%%%%%%%%%%%%%%%%%%%%%%%%%%%%%%%%%%%%%%%%%%%%%%%%%%%%%%%%%%%%%%%%

\section{Further insights on systematic mitigation}
\label{sec:insights_on_systematics}

%%%%%%%%%%%%%%%%%%%%%%%%%%%%%%%%%%%%%%%%%%%%%%%%%%%%%%%%%%%%%%%%%%%%%%%%%%%%%%%%%%%%%%%%%%%%%%%%%%%%%%%%%%%%%%%%%%%%%%%%%%%%%%%%%%%%

\subsection{Towards the ``futuristic'' limit}

In this work, we adopted three benchmark values for the number of GW events per year to test the scientific potential of GW clustering.
Therefore, it is quite natural to wonder whether these numbers can, in fact, be within reach or not.
In reality, the most important metric is given by the total number of GW events available for the clustering analysis, since the shot-noise scales proportionally to~$N^{-1/2}_\mathrm{GW}$.
In this sense, the number of events per year is completely degenerate with the total observation time; therefore, our results can always be recast in terms of running the GW observatory for a longer period of time.
Additionally, for the purpose of constraining PnG, BBHs are fundamentally equivalent to black hole-neutron star (BHNS) and neutron star (BNS) binaries.
Therefore, the number of GW events available for this kind of analysis is in fact larger than what is used for the forecasts, or, said otherwise, it is easier to reach a given number of total events by integrating in the analysis other kinds of compact object binaries.

The combination of current experimental sensitivity and large theoretical uncertainties in binary formation mechanisms makes it difficult to estimate the number of BNS and BHNS events available in the future.
As a primary example of this uncertainty, the reader can observe the different estimates of local merger rates for these classes of events obtained in two subsequent versions of Ligo-Virgo-KAGRA transient catalog, GWTC-4.0~\cite{abac:gwtc4} and GWTC-5.0~\cite{abac:gwtc5}.
Sometimes, even the same catalog analyzed with different numerical tools produces different estimates.
Therefore, here we provide some back-of-the-envelope calculation of how many additional events might be available for a PnG analysis, according to the latest estimates.

Regarding BNS, we have a local merger rate of~$R_{0,\mathrm{BNS}} \approx 2R_{0,\mathrm{BBH}}$; thus, it is reasonable to estimate~$N_\mathrm{BNS} \approx 2N_\mathrm{BBH}$, of which only~$40-50\%$ of them are detectable~\cite{iacovelli:3gdetectors}.
Similarly, for BHNS we have~$R_{0,\mathrm{BHNS}} \approx (1/2)R_{0,\mathrm{BBH}}$; therefore,~$N_\mathrm{BHNS} \approx (1/2)N_\mathrm{BBH}$, of which~$60-70\%$ should be detectable~\cite{iacovelli:3gdetectors}.
In the end, the total number of available GW events is approximately given by~$N^\mathrm{tot}_\mathrm{GW} \approx 2.3 N_\mathrm{BBH}$ or, stated otherwise, the same sensitivity can be achieved with half the total observation time.

Incorporating different populations in the same clustering analysis is certainly not so straightforward, since the BNS and BHNS events have typical radial and angular uncertainties that are larger than those of BBHs.
For example, the fraction of detected events with~$\Delta\Omega_{68} \lesssim 50\ \mathrm{deg}^2$ is~$90, 77, 70\%$ for BBH, BHNS, and BNS, respectively~\cite{iacovelli:3gdetectors}.\footnote{
In terms of Fisher forecasts, we have~$\Delta\Omega_{68} = (\Delta\chi^2_{68}/\Delta\chi^2_{90}) \Delta\Omega_{90} \simeq 0.5 \Delta\Omega_{90}$, since~$\Delta\chi^2_{68} = 2.33$ and~$\Delta\chi^2_{90}=4.61$.}
Combining maps with different angular and radial resolution requires a dedicated work on its own, and thus it is left for future work.
Additionally, by including additional populations the number of bias parameters naturally grows, which can potentially degrade the constraining power of GWxLSS cross-correlations.
However, especially in the case where binaries are all formed by the isolated binary channel, we could easily be in a situation where different categories of GW events act as different tracers of the same star-formation process.
Therefore, their biases will not be independent of each other.
In the end, including additional populations has its own challenges and advantages; however, these considerations show that the optimistic and futuristic limits are actually closer than what originally thought.

%%%%%%%%%%%%%%%%%%%%%%%%%%%%%%%%%%%%%%%%%%%%%%%%%%%%%%%%%%%%%%%%%%%%%%%%%%%%%%%%%%%%%%%%%%%%%%%%%%%%%%%%%%%%%%%%%%%%%%%%%%%%%%%%%%%%

\subsection{Degeneracies with assembly bias}

It is well-known that the local PnG~$f_\mathrm{NL}$ parameter is strongly degenerate with the amplitude of the assembly bias~$b_\phi$~\cite{barreira:bphifnlI, barreira:bphifnlII, barreira:bphifnlIII, lazeyras:bphifnl}.
Additionally, the true value of the assembly bias also has large uncertainties because of its deviation from universal behavior.
Numerical simulations suggest that the value of the parameter~$p$ varies at least between~$p_\mathrm{young} \approx -1.4$ and~$p_\mathrm{old} \approx 1.6$ for young and old halos, respectively, which led the community to look for new strategies to mitigate this uncertainty, see, e.g., Ref.~\cite{fondi:bphifnl}.

In this work, we did not actively try to place any constraints on the specific value of the assembly bias, since our focus was different.
Additionally, to our knowledge, no author has ever investigated or estimated the assembly bias of GW events.
In terms of analytical estimates of its magnitude, simple HOD considerations suggest that
\begin{equation}
    b_{\phi,\mathrm{GW}}(z) = n^{-1}_\mathrm{GW}(z) \int d\bm{\theta} \frac{dn_\mathrm{GW}}{d\bm{\theta}} b_\phi (\bm{\theta},z),
\end{equation}
where~$n_\mathrm{GW}$ is the GW number density, and~$\bm{\theta}$ represents the set of parameters that the assembly bias depends on.
Unfortunately, without running cosmological simulation with hydrodynamics, it turns out to be basically impossible to estimate the number of GW events depending on the relevant assembly bias parameters,~$dn_\mathrm{GW}/d\bm{\theta}$.
For instance, although a link between assembly bias and halo concentration has already been demonstrated, the actual concentration of individual halos deviates significantly with respect to the cosmological median value~\cite{wang:haloconcentration}.
Similarly, properties of central and satellite galaxies are tightly connected, see, e.g., Ref.~\cite{hearing:galacticconformity}; therefore, it becomes challenging to assess galaxy properties - as the star-formation rate - from global probability distribution functions.

However, our analysis allows us to provide a back-of-the-envelope estimate of what should be the precision we understand deviations from the universal behavior to avoid introducing statistical biases.
In fact, the reader can easily imagine how choosing an incorrect value of~$p$ is formally equivalent to introducing a non-zero value of~$b_\mathrm{sys}$.
In this sense, we have, at fixed~$f_\mathrm{NL}$, that the tolerable uncertainty on~$p$ reads as
\begin{equation}
    \sigma_p \approx \frac{\mathcal{T}_m(k,z) \left( k/k_\mathrm{sys}\right)^{\alpha_\mathrm{sys}}}{2\delta_c f_\mathrm{NL}} \sigma_{b_\mathrm{sys}|f_\mathrm{NL}},
\end{equation}
where~$\sigma_{b_\mathrm{sys}|f_\mathrm{NL}}= \sigma_{b_\mathrm{sys}} \sqrt{1-\rho^2_{b_\mathrm{sys}f_\mathrm{NL}}}$ is the conditional error on the systematic bias amplitude given a fixed value of~$f_\mathrm{NL}$, and~$\rho^2_{\alpha\beta}$ is the correlation coefficient between the~$\theta_\alpha$ and~$\theta_\beta$ parameters.
For~$f_\mathrm{NL}=1$, i.e., for the PNG fiducial model, we have~$\sigma_p \approx 1.75, 1.69, 1.35$ for the galaxy-only, conservative and futuristic GWxLSS scenarios, respectively.
Since in all cases~$\sigma_p \lesssim |p_\mathrm{old}-p_\mathrm{young}|$, it then becomes evidently clear how the control of assembly bias becomes critical in avoiding confusion between systematic and New Physics effects.
It also motivates further studies of the GW assembly bias, given their cross-correlation scientific potential.

%%%%%%%%%%%%%%%%%%%%%%%%%%%%%%%%%%%%%%%%%%%%%%%%%%%%%%%%%%%%%%%%%%%%%%%%%%%%%%%%%%%%%%%%%%%%%%%%%%%%%%%%%%%%%%%%%%%%%%%%%%%%%%%%%%%%

\section{Conclusions}
\label{sec:conclusions}

In addition to providing critical information about stellar evolution, GWs act also as LSS tracer and return a complementary picture of how structure formation unfolded.
In this work, we focussed on the latter aspect and we showcased how GW clustering help shedding light into one of the big open questions of modern Cosmology: how much Gaussian the primordial Universe was?
In particular, we showed how GW can not only provide tighter constraints on local PnG, but also mitigate systematic errors that appear at large scales and bias the statistical inference.
The first aspect was also analyzed in Ref.~\cite{gagnon:gwxlsspngconstraints}, which found~$\sigma_{f_\mathrm{NL}}=8.5$ for~$10^6\ \mathrm{GW/yr}$ due to the limited sky coverage of the Vera Rubin observatory.
In this sense, our work highlights how the combination with all-sky galaxy survey provides errors that are approximately five times smaller than what was previously reported for the same number of GW events.

On the other hand, the aspect connected to the control of systematics is radically new and exemplifies how accuracy can also be improved by the means of multi-tracer analysis.
We showed with our formalism, which can be easily connected to known sources of systematics, such as wrong choices of assembly bias or star contamination, that GWxLSS cross-correlations can mitigate the impact of both known and unknown systematics independently on their exact form, the magnitude of PnG effects, and the GW binning strategy.
In particular, for PnG models with~$f_\mathrm{NL} \gtrsim 2$, we show how cross-correlation can effectively rule-out a single-field slow-roll interpretation of the signal even systematics are introduced in the analysis.
The significance of these numbers could have been further increased by including in the analysis also external observations, such as CMB temperature, anisotropy, and lensing anisotropies.
We chose not to do so for two reasons.
First, we decided to present a Late-Universe-only analysis to avoid entering into the discussion of what happens when measurements apparently in tension are combined together, as in this current historical moment.
Second, in our estimation, adding the constraining power of the same dataset to all combination of Late-Universe probes would have just changed the quantitative aspect of the message, for instance by providing even smaller errors than what currently reported, but not the qualitative point we are making, which represents the true novelty.

Cross-correlations are not the only alternative to enhance sensitivity to specific cosmological parameters; for instance, it has been shown that it is possible to optimize the galaxy dataset analysis to specifically tighten constraints on~$f_\mathrm{NL}$, see, e.g., Refs.~\cite{slosar:optimizingfnl, barreira:optimizingfnl, karagiannis:optimizingfnl}.
Adding bispectrum information also increases the constraining power, both directly and indirectly by breaking degeneracies with cosmological and bias parameters.
Although both techniques turn out to be quite good in increasing precision, little is currently known about their capabilities in limiting the impact of systematics. 
In this sense, GWxLSS cross-correlation techniques keep their appeal in terms of mitigating the impact of systematics.
This idea is general enough that it works even if the GW clustering suffers from its own set of systematic effects, as long as they are not shared with galaxies.
We leave the analysis of this scenario for future work.

An interesting aspect to analyze is whether only BBHs of astrophysical origin act as a ``PnG detector''.
Previously we already discussed that other astrophysical objects, like BHNS and BNS, are also informative, although their localization is known less accurately than black hole binaries'.
However, in recent years, the community started to take a new interest in Primordial Black Holes, i.e., a class of black holes that formed in the Early Universe and whose formation process has nothing to do with stellar evolution processes~\cite{zeldovich:pbhformation, hawking:pbhformation, carr:pbhformation, chapline:pbhformation}.
Although their existence and abundance are a matter of current debate in the literature, we can already note that in this context it does not matter whether the origin of the BBH is astrophysical or primordial.
Even in the latter case they will behave as a tracer of LSS, despite primordial black hole binary formation mechanisms being radically different from those of astrophysical black holes.
Therefore, their bias is representative of a different kind of environments, but still carries the imprint of the matter field and displays the effects of PnG.
This sort of statement is even more radical: even in the case of binaries made by Exotic Compact Object, which appear in multiple extensions of General Relativity~\cite{cardoso:exoticcompactobkects}, it is very likely to develop a connection between a certain type of environment and the formation of compact object binaries; therefore, even these objects perfectly serve the purpose of controlling systematics for PnG detections.

In summary, the frontiers of GW astronomy promise to be even more astonishing than what originally thought, and the key to unlocking the true potential hidden in GW datasets remains creativity in looking for alternative fashions of analyzing data.
This work presents another simple, yet crucial, aspect of GW clustering, and shows how even a number of GW not comparable to the number of observed galaxies still allows us to enlarge and make more robust our knowledge on the formation and evolution of our Universe.

%%%%%%%%%%%%%%%%%%%%%%%%%%%%%%%%%%%%%%%%%%%%%%%%%%%%%%%%%%%%%%%%%%%%%%%%%%%%%%%%%%%%%%%%%%%%%%%%%%%%%%%%%%%%%%%%%%%%%%%%%%%%%%%%%%%%

\acknowledgments

NB acknowledges support from the European Union's Horizon Europe research and innovation program under the Marie Sk\l{}odowska-Curie grant agreement no. 101207487 (GWSKY - Mapping the Universe with Gravitational Waves).
AR acknowledges funding from the Italian Ministry of University and Research (MIUR) through the ``Dipartimenti di eccellenza'' project ``Science of the Universe''.

%%%%%%%%%%%%%%%%%%%%%%%%%%%%%%%%%%%%%%%%%%%%%%%%%%%%%%%%%%%%%%%%%%%%%%%%%%%%%%%%%%%%%%%%%%%%%%%%%%%%%%%%%%%%%%%%%%%%%%%%%%%%%%%%%%%%

\appendix
\include{Appendix}

%%%%%%%%%%%%%%%%%%%%%%%%%%%%%%%%%%%%%%%%%%%%%%%%%%%%%%%%%%%%%%%%%%%%%%%%%%%%%%%%%%%%%%%%%%%%%%%%%%%%%%%%%%%%%%%%%%%%%%%%%%%%%%%%%%%%

\bibliography{bibliography}
\bibliographystyle{utcaps}

\end{document}

%% file: Appendix.tex
\section{Properties of large-scale structure tracers}
\label{app:lss_tracers}

\begin{table}[ht]
    \centerline{
    \begin{tabular}{|c|cccc|}
        \hline
        Galaxy sample & \# bins & $\ell_{\min}$ & $\ell_{\max}$ & Redshift mean~$z_j$ \\
        \hline
        \hline
        Large Scales & 10 & 2    & 100  & 0.4, 0.8, 1.2, 1.6, 2.0, 2.4, 2.8, 3.2, 3.6, 4.0 \\
        \hline
        \hline
        \multirow{5}{*}{Small Scales} & 10 & 101    & 250  & 0.4, 0.8, 1.2, 1.6, 2.0, 2.4, 2.8, 3.2, 3.6, 4.0 \\
         & 9  & 251  & 650  & 0.8, 1.2, 1.6, 2.0, 2.4, 2.8, 3.2, 3.6, 4.0 \\
         & 8  & 651  & 1250 & 1.2, 1.6, 2.0, 2.4, 2.8, 3.2, 3.6, 4.0 \\
         & 7  & 1251 & 2300 & 1.6, 2.0, 2.4, 2.8, 3.2, 3.6, 4.0 \\
         & 6  & 2301 & 3000 & 2.0, 2.4, 2.8, 3.2, 3.6, 4.0 \\
        \hline
    \end{tabular}}
    \vspace{0.3cm}
    \centerline{
    \begin{tabular}{|c|cccc|}
        \hline
        GW sample & \# bins & $\ell_{\min}$ & $\ell_{\max}$ & Redshift mean~$z_j$ \\
        \hline
        \hline
        Large Scales & 8 & 2 & 100 & 0.4, 0.8, 1.2, 1.6, 2., 2.4, 2.9, 3.6 \\
        \hline
    \end{tabular}}
    \caption{\textit{Top table}: SPHEREx-like galaxy multipole ranges and corresponding redshift binning.
    \textit{Bottom table}: GW population multipole ranges and corresponding redshift binning.}
    \label{tab:tracers_binning_properties}
\end{table}

%%%%%%%%%%%%%%%%%%%%%%%%%%%%%%%%%%%%%%%%%%%%%%%%%%%%%%%%%%%%%%%%%%

\subsection{Galaxy sample}
\label{subapp:galaxy_sample}

We define a strawman distribution of SPHEREx-like galaxies~\cite{dore:spherexwhitepaperI, dore:spherexwhitepaperII, dore:spherexwhitepaperIII}, with ten redshift bins in the redshift interval~$[0.2,4.2]$, with redshift mean~$z_j$ reported in table~\ref{tab:tracers_binning_properties}, and equal half-width~$\Delta z=0.2$.
We use a top-hat window function for all redshift bins.
Following Ref.~\cite{bosi:gwxlssI}, we parametrize the galaxy distribution as
\begin{equation}
    \frac{d^2 N_\mathrm{gal}}{dzd\Omega} = \mathcal{A}_\mathrm{gal} \left(\frac{z}{z_\mathrm{gal}}\right)^{\alpha_\mathrm{gal}} e^{-\left(z/z_\mathrm{gal}\right)^{\beta_\mathrm{gal}}},
\end{equation}
where~$\{\mathcal{A}_\mathrm{gal}, z_\mathrm{gal}, \alpha_\mathrm{gal}, \beta_\mathrm{gal} \}=\{25509\ \mathrm{gal/deg^2}, 0.09, 1.75, 0.69\}$.
The Gaussian component of the galaxy bias is given by
\begin{equation}
    b_\mathrm{gal} = b_\mathrm{0,gal} + b_\mathrm{1,gal}z + b_\mathrm{2,gal} z^2,
\end{equation}
where~$\{ b_{0,\mathrm{gal}}, b_{1,\mathrm{gal}}, b_{2,\mathrm{gal}} \}=\{0.53, 1.59, -0.08\}$, while the magnification bias is chosen to be~$s_\mathrm{gal} = 0.6$, and the evolution bias is computed directly from the galaxy redshift distribution.
In the case of our galaxy sample, shot noise is the only noise contribution for the scales considered in this work, and it is modeled as
\begin{equation}
    N^\mathrm{gal,gal}_\ell (z_i, z_j) = \left( \frac{dN_\mathrm{gal}}{d\Omega} \right)^{-1} \delta_{ij}^K.
\end{equation}

In our analysis, we consider only clustering at linear scales; thus, we impose a cut in terms of maximum multipoles.
Following Ref.~\cite{bellomo:multiclass}, the maximum multipole in each redshift bin is computed as
\begin{equation}
    \ell_{\mathrm{max,j}}\simeq k_{\mathrm{max}}(z_{j})r(z_{j}),
\end{equation}
where~$k_{\mathrm{max}}(z_{j})$ is the non-linearity scale at redshift~$z_j$, and~$r(z_j)$ is its comoving distance.
The non-linearity scale is usually computed as the scale where the variance of fluctuations of the smoothed matter field becomes order unity.
In other words, we need to estimate at which scale~$R_\mathrm{max} = k_\mathrm{max}^{-1}$ such that~$\sigma(R_{\mathrm{max}},z)=1$, where the variance is explicitly defined as
\begin{equation}
    \sigma^{2}(R,z)=\int\frac{d^{3}k}{(2\pi)^{3}}W_{R}^{2}(k,R) P_m(k,z),
\end{equation}
the matter field power spectrum is indicated by~$P_m$, and the filter function~$W_R$ is usually assumed to be a top-hat in real space:
\begin{equation}
    W_R(k,R) = 3\frac{\sin(kR) - kR\cos(kR)}{(kR)^3}.
\end{equation}
The resulting maximum multipoles are also reported in table~\ref{tab:tracers_binning_properties}.
For redshift~$z \geq 2$ the allowed multipole would be larger than~$\ell_{\mathrm{max}}=3000$; however, we enforce this upper bound to save computational time. 
The core message of the article does not change when including multipoles~$\ell \geq 3000$.

%%%%%%%%%%%%%%%%%%%%%%%%%%%%%%%%%%%%%%%%%%%%%%%%%%%%%%%%%%%%%%%%%%%%%%%%%%%%%%%%%%%%

\subsection{Gravitational wave sample}
\label{subapp:gw_sample}

The second LSS tracer in our analysis is a population of resolved GWs of astrophysical origin, as explained in Ref.~\cite{bosi:gwxlssI}.
In summary, we parametrize the GW redshift distribution as
\begin{equation}
    \frac{d^2 N_\mathrm{GW}}{dzd\Omega} = \mathcal{A}_\mathrm{GW} \left(\frac{z}{z_\mathrm{GW}}\right)^{\alpha_\mathrm{GW}} e^{-\left(z/z_\mathrm{GW}\right)^{\beta_\mathrm{GW}}},
\end{equation}
where~$\{\mathcal{A}_\mathrm{GW}, z_\mathrm{GW}, \alpha_\mathrm{GW}, \beta_\mathrm{GW} \} = \{9.138 \times 10^{-11}\ \mathrm{GW/deg^2}, 2.02\times 10^{-3}, 6.12, 0.41\}$.
These values assume a third-generation detector network given by Einstein Telescope and two Cosmic Explorer. 
The amplitude has been rescaled from the original value to predict a total number of~$10^5$ GW events per year in the conservative scenario.
The Gaussian component of the GW bias is modeled as
\begin{equation}
    b_\mathrm{GW}(z) = b_{0,\mathrm{GW}} + b_{1,\mathrm{GW}} z + b_{2, \mathrm{GW}} z^2,
\end{equation}
where~$\{ b_{0,\mathrm{GW}}, b_{1,\mathrm{GW}}, b_{2,\mathrm{GW}} \} = \{ 0.65, 1.45, -0.14 \}$, and we discard the term~$b_{3,\mathrm{GW}} z^3$ since it affects the bias only at the percent level.
The magnification bias is set to~$s_\mathrm{GW} = 0$.

Since the uncertainty on the radial distance of GW events is significantly larger than that of galaxies, we introduce for this sample a Gaussian window function.
We consider a baseline number of eight redshift bins with half-width never smaller than the redshift uncertainty, as also explained in appendix~C of Ref.~\cite{bosi:gwxlssI}.
The values for the redshift mean and half-width are reported in table~\ref{tab:tracers_binning_properties}.
Similarly, GWs also have significantly poorer spatial resolution than that of galaxy surveys.
Therefore, the number of accessible multipoles is drastically reduced and ultimately limited to~$\ell \lesssim 180^{\circ}/\theta_\mathrm{res}^\mathrm{max}$, where~$\theta_\mathrm{res}^\mathrm{max}$ is the maximum resolution of the detector network. 

The noise contribution for the GW sample takes into account for both these limitations, as already explained in Ref.~\cite{bosi:gwxlssI}.
To account for uncertainties in the radial distribution, we define the observed distribution
\begin{equation}
    \frac{dN_\mathrm{GW}^\mathrm{obs}}{d\Omega} = \int_{z_\mathrm{min}}^{z_\mathrm{max}} dz_\mathrm{obs} \int_0^\infty dz_\mathrm{GW} \frac{dN_\mathrm{GW}}{dzd\Omega}(z_\mathrm{GW}) p_\mathrm{obs} (z_\mathrm{GW} ,  z_\mathrm{obs}),
\end{equation}
where~$p_\mathrm{obs} (z_\mathrm{GW},  z_\mathrm{obs})$ is the probability of observing at redshift~$z_{\text{obs}}$ a GW emitted in reality at redshift~$z_{\text{GW}}$ due to the noisiness of the measurement, which we assume to be
\begin{equation}
    p_\mathrm{obs}(z_\mathrm{GW}, z_\mathrm{obs}) = \frac{1}{\sqrt{2\pi\sigma_z^2}} \exp \left[ - \frac{(z_\mathrm{GW} - z_\mathrm{obs})^2}{2\sigma_z^2} \right].
\end{equation}
On the other hand, the effect of poor angular resolution is included as a beam smearing term.
In the end, the noise reads as
\begin{equation}
    N_\ell^\mathrm{GW,GW}(z_i, z_j) = \left( \frac{dN_\mathrm{GW}^\mathrm{obs}}{d\Omega} \right)^{-1} \exp\left[ \frac{\ell (\ell+1)(\theta_\mathrm{res}^\mathrm{avg})^2}{8\log2} \right] \delta^K_{ij}.
\end{equation}

%%%%%%%%%%%%%%%%%%%%%%%%%%%%%%%%%%%%%%%%%%%%%%%%%%%%%%%%%%%%%%%%%%%%%%%%%%%%%%%%%%%%%%%%%%%%%%%%%%%%%%%%%%%%%%%%%%%%%%%%%%%%%%%%%%%%

\section{Fisher forecasts}
\label{app:fisher_forecast}

\begin{table}[ht]
    \begin{center}
        PNG~$\left( \alpha_\mathrm{sys}=-2 \right)$ 
    \end{center}
    \vspace*{-0.25cm}
    \centerline{
    \begin{tabular}{|c|c|c|c|c|c|c|c|c|c|c|}
    \hline
    $\omega_\mathrm{cdm}$ & $h$ & $n_s$ & $b_{0,\mathrm{gal}}$ & $b_{1,\mathrm{gal}}$ & $b_{2,\mathrm{gal}}$ & $b_{0,\mathrm{GW}}$ & $b_{1,\mathrm{GW}}$ & $b_{2,\mathrm{GW}}$ & $f_\mathrm{NL}$ & $b_\mathrm{sys}$ \\
    \hline
    \hline
    0.1202 & 0.67556 & 0.9649 & 0.53 & 1.59 & -0.08 & 0.65 & 1.45 & -0.14 & 1 & 0 \\
    \hline
    \end{tabular}}
    \vspace*{0.2cm}
    \begin{center}
        SYS~$\left( \alpha_\mathrm{sys}=-2 \right)$     
    \end{center}
    \vspace*{-0.25cm}
    \centerline{
    \begin{tabular}{|c|c|c|c|c|c|c|c|c|c|c|}
    \hline
    $\omega_\mathrm{cdm}$ & $h$ & $n_s$ & $b_{0,\mathrm{gal}}$ & $b_{1,\mathrm{gal}}$ & $b_{2,\mathrm{gal}}$ & $b_{0,\mathrm{GW}}$ & $b_{1,\mathrm{GW}}$ & $b_{2,\mathrm{GW}}$ & $f_\mathrm{NL}$ & $b_\mathrm{sys}$ \\
    \hline
    \hline
    0.1202 & 0.67556 & 0.9649 & 0.53 & 1.59 & -0.08 & 0.65 & 1.45 & -0.14 & 0 & 0.0002 \\
    \hline
    \end{tabular}}
    \caption{Fiducial value for the cosmological, bias, PnG, and systematic parameters in the PNG and SYS fiducial models.}
    \label{tab:fisher_fiducial_values}
\end{table}

\begin{figure}[ht]
    \centerline{
    \includegraphics[width=\linewidth]{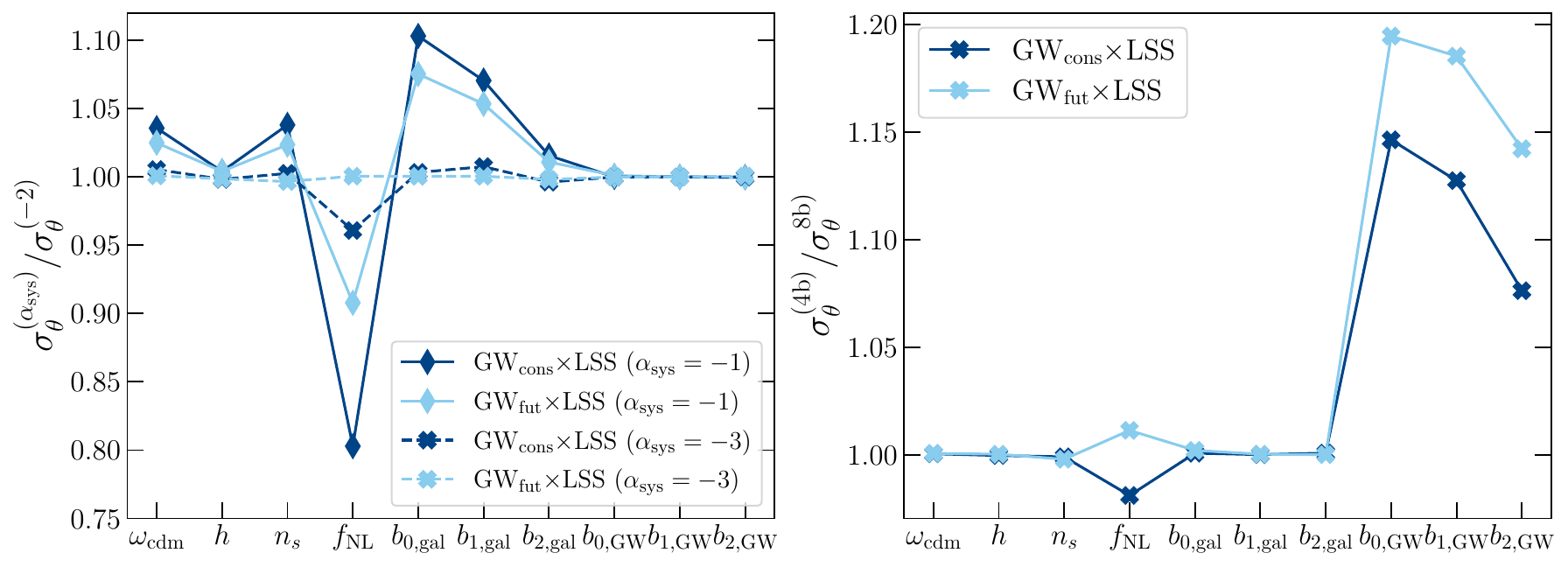}}
    \caption{\textit{Left panel}: ratio of marginalized errors between the~$\alpha_\mathrm{sys} = \{-1, -3\}$ and~$\alpha_\mathrm{sys}=-2$ GW~$\times$LSS PNG/sys analysis.
    \textit{Right panel}: ratio of marginalized errors between the binning scheme involving four and eight redshift bins for the GW population. 
    Ratios are reported for the GW~$\times$LSS PNG/sys analysis.}
    \label{fig:asys_4bins_params}
\end{figure}

We report in table~\ref{tab:fisher_fiducial_values} the fiducial values for the cosmological, bias, PnG, and systematic parameters for the PNG and SYS models.
Tables~\ref{tab:PNG_errors} and~\ref{tab:SYS_errors} report the marginalized errors on the cosmological, bias, PnG, and systematic parameters for different combinations of tracers and fiducial models.
Figure~\ref{fig:asys_4bins_params} shows the ratio of marginalized errors in scenarios involving different systematic tilts and GW rebinning schemes.

\begin{table}[ht]
    \centerline{
    \begin{tabular}{|c||c|c||c|c||c|c|}
    \hline
    Gal-only & LS/no-sys& LS/sys & SS/no-sys & SS/sys & LS+SS/no-sys & LS+SS/sys \\
    \hline
    \hline
    $\sigma_{\omega_\mathrm{cdm}}$ & 0.03 & 0.03 & 0.0031 & 0.0032 & 0.0027 & 0.0027 \\
    $\sigma_h$ & 0.15 & 0.15 & 0.011 & 0.011 & 0.011 & 0.011 \\
    $\sigma_{n_s}$ & 0.05& 0.05& 0.0067 & 0.0071 & 0.0058 & 0.0058 \\
    \hline
    \hline
    $\sigma_{b_{0g}}$ & 0.09& 0.09 & 0.01 & 0.01 & 0.01 & 0.01 \\
    $\sigma_{b_{1g}}$ & 0.16 & 0.16 & 0.02 & 0.02 & 0.02 & 0.02 \\
    $\sigma_{b_{2g}}$ & 0.01 & 0.01 & 0.03 & 0.003 & 0.003 & 0.003 \\
    \hline
    \hline
    $\sigma_{f_\mathrm{NL}}$ & 1.81 & 2.23& 7.09 & 9.56& 1.24 & 1.72 \\
    $10^{4}\sigma_{b_\mathrm{sys}}$ & - & 3.6 & - & 6.2 & - & 3.1 \\
    \hline
    \end{tabular}}
    \vspace{0.2cm}
    \centerline{
    \begin{tabular}{|c||c|c||c|c||c|c|}
    \hline
    Gal-GW & Cons/no-sys & Cons/sys & Opt/no-sys & Opt/sys  & Fut/no-sys & Fut/sys \\
    \hline
    \hline
    $\sigma_{\omega_\mathrm{cdm}}$  & 0.03 (0.0027)& 0.03 (0.0027)& 0.02 (0.0027)& 0.02 (0.0027)&0.02 (0.0026)&0.016 (0.0026)\\
    $\sigma_h$  & 0.14 (0.011)& 0.14 (0.011)& 0.11 (0.011)& 0.11 (0.011)&0.087 (0.011)&0.087 (0.011)\\
    $\sigma_{n_s}$ & 0.04 (0.0058)& 0.04 (0.0058)& 0.03 (0.0056)& 0.03(0.0056)&0.02 (0.0052)& 0.02 (0.0053)\\
    \hline
    \hline
    $\sigma_{b_{0, \mathrm{gal}}}$  & 0.08 (0.008)& 0.08 (0.008)& 0.07 (0.008)& 0.07 (0.008)&0.05 (0.008)& 0.05 (0.008)\\
    $\sigma_{b_{1, \mathrm{gal}}}$  & 0.14 (0.02)& 0.14 (0.02)& 0.11 (0.02)& 0.11 (0.02)& 0.09 (0.02)&0.09 (0.02)\\
    $\sigma_{b_{2, \mathrm{gal}}}$ & 0.01 (0.003)& 0.01 (0.003)& 0.01 (0.003)& 0.01 (0.003)&0.01 (0.003)& 0.01 (0.003)\\
    $\sigma_{b_{0, \mathrm{GW}}}$ & 0.23 (0.21)& 0.23 (0.21)& 0.10 (0.07)& 0.10 (0.07)&0.07 (0.04)& 0.07 (0.04)\\
    $\sigma_{b_{1, \mathrm{GW}}}$  & 0.34 (0.32)& 0.34 (0.32)& 0.13 (0.10)& 0.14 (0.10)&0.09 (0.06)& 0.09 (0.06)\\
    $\sigma_{b_{2, \mathrm{GW}}}$  & 0.10 (0.10)& 0.10 (0.10)& 0.03 (0.03)& 0.03 (0.03)&0.02 (0.02)& 0.02 (0.02)\\
    \hline
    \hline
    $\sigma_{f_\mathrm{NL}}$ & 1.75 (1.22)& 2.14 (1.67)& 1.47 (1.09)& 1.69 (1.36)&1.22 (0.93)& 1.32 (1.07)\\
    $10^{4}\sigma_{b_\mathrm{sys}}$ & - & 3.5 (3.0)& - & 3.0 (2.7)&-&2.6 (2.4)\\
    \hline
    \end{tabular}}
    \caption{Marginalized errors for the PNG fiducial model, for both PNG/no-sys and PNG/sys subclasses of the analysis.
    \textit{Top table}: galaxy-only marginalized errors with all combinations of scales.
    \textit{Bottom table}: GWxLSS marginalized errors at large scales for a conservative, optimistic, and futuristic number of observed GW events.
    We report in parenthesis the marginalized errors obtained with the inclusion of the galaxy small-scales.
    These errors refer to the~$\alpha_\mathrm{sys}=-2$ scenario.}
    \label{tab:PNG_errors}
\end{table}

\begin{table}[ht]
    \centerline{
    \begin{tabular}{|c||c|c||c|c||c|c|}
    \hline
    Gal-only & LS/no-png & LS/png & SS/no-png & SS/png & LS+SS/no-png  & LS+SS/png \\
    \hline
    \hline
    $\sigma_{\omega_\mathrm{cdm}}$ & 0.03& 0.03& 0.0032& 0.0032& 0.0027& 0.0027\\
    $\sigma_h$ & 0.15& 0.15& 0.011& 0.011& 0.011& 0.011\\
    $\sigma_{n_s}$ & 0.047& 0.049& 0.0071& 0.0071& 0.0058& 0.0058\\
    \hline
    \hline
    $\sigma_{b_{0, \mathrm{gal}}}$ & 0.09& 0.09& 0.009& 0.009& 0.008& 0.008\\
    $\sigma_{b_{1, \mathrm{gal}}}$ & 0.16& 0.16& 0.02& 0.02& 0.02& 0.02\\
    $\sigma_{b_{2, \mathrm{gal}}}$ & 0.01& 0.01& 0.003& 0.003& 0.003& 0.003\\
    \hline
    \hline
    $\sigma_{f_\mathrm{NL}}$ & - & 2.21& - & 9.56& - & 1.75\\
    $10^{4}\sigma_{b_\mathrm{sys}}$ & 3.1& 3.7& 47& 63& 2.3& 3.2\\
    \hline
    \end{tabular}}
    \vspace{0.2cm}
    \centerline{
    \begin{tabular}{|c||c|c||c|c||c|c|}
    \hline
    Gal-GW & Cons/no-png & Cons/png & Opt/no-png & Opt/png & Fut/no-png & Fut/png \\
    \hline
    \hline
    $\sigma_{\omega_\mathrm{cdm}}$ & 0.03 (0.0027)& 0.03 (0.0027)& 0.02 (0.0026)& 0.02 (0.0027)& 0.02 (0.0026)& 0.02 (0.0026)\\
    $\sigma_h$ &  0.14 (0.011)& 0.14 (0.011)& 0.11 (0.011)& 0.11 (0.011)&  0.084 (0.011)& 0.088 (0.011)\\
    $\sigma_{n_s}$ & 0.04 (0.0058)& 0.04 (0.0058)& 0.03 (0.0056)& 0.03 (0.0056)& 0.02 (0.0053)& 0.02 (0.0053)\\
    \hline
    \hline
    $\sigma_{b_{0, \mathrm{gal}}}$ & 0.08 (0.008)& 0.08 (0.008)&  0.06 (0.008)& 0.07 (0.008)& 0.05 (0.008)& 0.05 (0.008)\\
    $\sigma_{b_{1,\mathrm{gal}}}$ & 0.14 (0.018)& 0.15 (0.018)& 0.11 (0.018)& 0.11 (0.018)& 0.08 (0.017)& 0.09 (0.017)\\
    $\sigma_{b_{2, \mathrm{gal}}}$ & 0.01 (0.003)& 0.01 (0.003)& 0.01 (0.003)& 0.01 (0.003)& 0.01 (0.003)& 0.01 (0.003)\\
    $\sigma_{b_{0, \mathrm{GW}}}$ & 0.23 (0.21)& 0.23 (0.21)& 0.10 (0.07)& 0.10 (0.07)& 0.07 (0.04)& 0.07 (0.04)\\
    $\sigma_{b_{1, \mathrm{GW}}}$ & 0.34 (0.32)& 0.34 (0.32)& 0.13 (0.10)& 0.14 (0.10)& 0.09 (0.06)& 0.09 (0.06)\\
    $\sigma_{b_{2, \mathrm{GW}}}$ & 0.10 (0.10)& 0.10 (0.10)& 0.03 (0.03)& 0.03 (0.03)& 0.02 (0.02)& 0.02 (0.02)\\
    \hline
    \hline
    $\sigma_{f_\mathrm{NL}}$ & - & 2.12 (1.70)& - & 1.65 (1.38)& - & 1.28 (1.08)\\
    $10^{4}\sigma_{b_\mathrm{sys}}$ & 3.0 (2.2)& 3.6 (3.1)& 2.8 (2.2)& 3.1 (2.8)&  2.5 (2.1)& 2.6 (2.4)\\
    \hline
    \end{tabular}}
    \caption{Marginalized errors for the SYS fiducial model, for both SYS/no-png and SYS/png subclasses of the analysis.
    \textit{Top table}: galaxy-only marginalized errors with all combinations of scales.
    \textit{Bottom table}: GWxLSS marginalized errors at large scales for a conservative, optimistic, and futuristic number of observed GW events.
    We report in parenthesis the marginalized errors obtained with the inclusion of the galaxy small-scales.
    These errors refer to the~$\alpha_\mathrm{sys}=-2$ scenario.}
    \label{tab:SYS_errors}
\end{table}

%%%%%%%%%%%%%%%%%%%%%%%%%%%%%%%%%%%%%%%%%%%%%%%%%%%%%%%%%%%%%%%%%%%%%%%%%%%%%%%%%%%%%%%%%%%%%%%%%%%%%%%%%%%%%%%%%%%%%%%%%%%%%%%%%%%%

\section{\texttt{Multi\_CLASS v2}}
\label{app:multiclass}

In this appendix, we describe the changes introduced in the new version of \texttt{Multi\_CLASS}. 
With respect to the previous version, the code now allows for the independent calculation of two classical LSS tracers and the resolved GW number count angular power spectra, each characterized by independent set of input values, namely redshift means and half-widths, window functions, clustering, assembly, magnification, and evolution bias functions. 
Resolved GW have been isolated in light of future applications, even though their underlying theory is the same as traditional LSS number counts.
The code now computes all of the spectra, including the cross-correlation angular power spectra between different tracers.
In other words, in a single run it can compute all possible combinations of observables, improving on the previous implementation where three runs were needed for the same task. 
Additionally, we improved the treatment of bias parameters that have now been promoted to redshift-dependent functions. 
This change alone removes a~$\mathcal{O}(1-10\%)$ computational error in the calculation of the~$C_\ell$'s.
Finally, we allow the user to specify the assembly bias in the case PnG are included in the computation.
These bias functions are hard-coded in the~\texttt{transfer.c} module.

%%%%%%%%%%%%%%%%%%%%%%%%%%%%%%%%%%%%%%%%%%%%%%%%%%%%%%%%%%%%%%%%%%%%%%%%%%%%%%%%%%%%%%%%%%%%%%%%%%%%%%%%%%%%%%%%%%%%%%%%%%%%%%%%%%%%

\subsection{\texttt{Multi\_CLASS v2} for users}

Due to the split between LSS and resolved GW cumber count angular power spectra, the user has to specify which one they request as~\texttt{[output = nCl, rgwCl]}. 
When both of them are included, the code will also compute all the cross-correlations between them.

%%%%%%%%%%%%%%%%%%%%%%%%%%%%%%%%%%%%%%%%%%%%%%%%%%%%%%%%%%%%%%%%%%%%%%%%%%%%%%%%%%%%%%%%%%%%%%%%%%%%%%%%%%%%%%%%%%%%%%%%%%%%%%%%%%%%

\subsubsection*{Traditional LSS tracers}

These options preserve the possibility of computing multi-tracer angular power spectra between two galaxy populations, for instance.
The input options are characterized by the subscripts \texttt{\_1} and \texttt{\_2}, where the latter are switched off by default, and read only if the option \texttt{[selection\_multitracing = yes]} is given.

\begin{itemize}
    \item \texttt{selection\_mean\_1}, \texttt{selection\_mean\_2}: allow the user to specify the list of mean redshifts~$z_i$ of each tracer. 
    The number of redshift bins of the two tracers does not need to match, the only condition is that~$N_1\geq N_2$.
    No default value is provided.
    
    \item \texttt{selection\_width\_1}, \texttt{selection\_width\_2}: allow the user to specify the list of redshift bin half-widths~$\Delta z_i$ of each tracer. 
    The length of each list must match that of the respective mean redshift list.
    No default value is provided.
    
    \item \texttt{selection\_window\_1}, \texttt{selection\_window\_2}: allow the user to specify the window functions~$W(z,z_{i},\Delta z)$ of each tracer. 
    As in the previous version, the options available are \texttt{[gaussian, tophat, dirac]}. 
    No default option is provided.

    \item \texttt{selection\_tracer\_1}, \texttt{selection\_tracer\_2}: allow the user to describe whether the tracer model is hard-coded or read from a file.
    The two options available are~\texttt{[analytic/file]}, no default value is provided.
    This option was significantly changed from the previous version.
    
    \item \texttt{selection\_dNdz\_1}, \texttt{selection\_dNdz\_2}: allow the user to specify the analytical model of~$dN_X/dz$.
    Read only if \texttt{analytic} is selected, no default value is provided.
    This option was significantly changed from the previous version.
    
    \item \texttt{selection\_bias\_1}, \texttt{selection\_bias\_2}:  allow the user to specify the analytical model of~$b_X(z)$.
    Read only if \texttt{analytic} is selected, no default value is provided.

    \item \texttt{selection\_magnification\_bias\_1}, \texttt{selection\_magnification\_bias\_2}: allow the user to specify the analytical model of~$s_X(z)$.
    Read only if \texttt{analytic} is selected, no default value is provided.
    
    \item \texttt{selection\_evolution\_bias\_1}, \texttt{selection\_evolution\_bias\_2}: allow the user to specify the analytical model of~$f_X^\mathrm{evo}(z)$.
    Read only if \texttt{analytic} is selected, no default value is provided.

    \item \texttt{selection\_assembly\_bias\_1}, \texttt{selection\_assembly\_bias\_2}: allow the user to specify the analytical model of~$b_{\phi,X}(z)$.
    Read only if \texttt{analytic} is selected and PnG are present, no default value is provided.

    \item \texttt{selection\_dNdz\_filepath\_1}, \texttt{selection\_dNdz\_filepath\_2}: path to files containing the linear bias for each individual tracer.
    The files should contain two columns,~$\left[ z, dN_X/dz(z) \right]$.
    Read only if \texttt{file} is selected, no default path is provided.
    
    \item \texttt{selection\_bias\_filepath\_1}, \texttt{selection\_bias\_filepath\_2}: path to files containing the linear bias for each individual tracer.
    The files should contain two columns,~$\left[ z, b_X(z) \right]$.
    Read only if \texttt{file} is selected, no default path is provided.

    \item \texttt{selection\_magnification\_bias\_filepath\_1}, \texttt{selection\_magnification\_bias\_filepath\_2}: path to files containing the linear bias for each individual tracer.
    The files should contain two columns,~$\left[ z, s_X(z) \right]$.
    Read only if \texttt{file} is selected, no default path is provided.

    \item \texttt{selection\_evolution\_bias\_filepath\_1}, \texttt{selection\_evolution\_bias\_filepath\_2}: path to files containing the linear bias for each individual tracer.
    The files should contain two columns,~$\left[ z, f^\mathrm{evo}_X(z) \right]$.
    Read only if \texttt{file} is selected, no default path is provided.

    \item \texttt{selection\_assembly\_bias\_filepath\_1}, \texttt{selection\_assembly\_bias\_filepath\_2}: path to files containing the linear bias for each individual tracer.
    The files should contain two columns,~$\left[ z, b_{\phi,X}(z) \right]$.
    Read only if \texttt{file} is selected and PnG are present, no default path is provided.    
\end{itemize}

The option~\texttt{non\_diagonal} has been completely removed. 
The code now automatically computes all off-diagonal terms.

%%%%%%%%%%%%%%%%%%%%%%%%%%%%%%%%%%%%%%%%%%%%%%%%%%%%%%%%%%%%%%%%%%%%%%%%%%%%%%%%%%%%%%%%%%%%%%%%%%%%%%%%%%%%%%%%%%%%%%%%%%%%%%%%%%%%

\subsubsection*{Resolved gravitational wave events}

The logic of the input commands for the resolved GW number count angular power spectra follows that of LSS tracers.
We report the options for completeness, the usage is the same of that described above.

\begin{itemize}
    \item \texttt{selection\_mean\_rgw}; 

    \item \texttt{selection\_width\_rgw};
        
    \item \texttt{selection\_window\_rgw};.

    \item \texttt{selection\_tracer\_rgw};

    \item \texttt{selection\_dNdz\_rgw}; 
    
    \item \texttt{selection\_bias\_rgw};

    \item \texttt{selection\_magnification\_bias\_rgw};
    
    \item \texttt{selection\_evolution\_bias\_rgw};
    
    \item \texttt{selection\_assembly\_bias\_rgw};

    \item \texttt{selection\_dNdz\_filepath\_rgw}; 
    
    \item \texttt{selection\_bias\_filepath\_rgw};

    \item \texttt{selection\_magnification\_bias\_filepath\_rgw};
    
    \item \texttt{selection\_evolution\_bias\_filepath\_rgw};
    
    \item \texttt{selection\_assembly\_bias\_filepath\_rgw};
\end{itemize}

%%%%%%%%%%%%%%%%%%%%%%%%%%%%%%%%%%%%%%%%%%%%%%%%%%%%%%%%%%%%%%%%%%%%%%%%%%%%%%%%%%%%%%%%%%%%%%%%%%%%%%%%%%%%%%%%%%%%%%%%%%%%%%%%%%%%

\subsubsection*{Output}

We also report the syntax of the output file for completeness.
\begin{itemize}
    \item The traditional LSS number count angular power spectra~$C_{l}^{\text{XY}}(z_{i}z_{j})$ are labeled as \\
    \texttt{dens[index\_tracer][index\_bin]}-\texttt{dens[index\_tracer][index\_bin]}, where \texttt{index\_tracer} refers to the LSS population.
    
    \item Resolved GW angular power spectra are labeled as~\texttt{rgw[index\_bin]}.
    
    \item Cross-tracer angular power spectra are labeled as \texttt{dens[index\_tracer][index\_bin]}-~\texttt{rgw[index\_bin]}. 

    \item If~\texttt{selection\_multitracing = no}, \texttt{[index\_tracer]} is removed and only \texttt{[index\_bin]} is displayed.
\end{itemize}

%%%%%%%%%%%%%%%%%%%%%%%%%%%%%%%%%%%%%%%%%%%%%%%%%%%%%%%%%%%%%%%%%%%%%%%%%%%%%%%%%%%%%%%%%%%%%%%%%%%%%%%%%%%%%%%%%%%%%%%%%%%%%%%%%%%%

\subsubsection*{Primordial non-Gaussianity}

The implementation of PnG in this version of \texttt{Multi\_CLASS} is the same of the previous one applied to both LSS traditional tracers and resolved GWs.
However, we added the possibility of specifying the assembly bias, instead of adopting the universal relation.

%%%%%%%%%%%%%%%%%%%%%%%%%%%%%%%%%%%%%%%%%%%%%%%%%%%%%%%%%%%%%%%%%%%%%%%%%%%%%%%%%%%%%%%%%%%%%%%%%%%%%%%%%%%%%%%%%%%%%%%%%%%%%%%%%%%%